\documentclass[aas_macros]{mnras}

\usepackage{graphicx}
\usepackage{color}
\usepackage{amssymb}
\usepackage{ulem}
\usepackage{courier}
\usepackage[T1]{fontenc}
\usepackage{ae,aecompl}

\date{}

\title[Delayed triggering of radio AGN]{Delayed triggering of radio Active Galactic Nuclei in gas-rich minor mergers in the local Universe}
\author[S. S. Shabala et al.]{S. S. Shabala$^{1}$\thanks{e-mail:Stanislav.Shabala@utas.edu.au}, A. Deller$^{2}$, S. Kaviraj$^{3}$, E. Middelberg$^{4}$, R. J. Turner$^{1}$, Y. S. Ting$^{5}$,
\newauthor J. R. Allison$^{6}$, T. A. Davis$^{3,7}$\\
$^{1}$ School of Physical Sciences, Private Bag 37, University of Tasmania, Hobart, TAS 7001, Australia \\
$^{2}$ The Netherlands Institute for Radio Astronomy (ASTRON), Dwingeloo, The Netherlands \\
$^{3}$ Centre for Astrophysics Research, University of Hertfordshire, College Lane, Hatfield, Herts AL10 9AB, UK \\
$^{4}$ Astronomisches Institut der Ruhr-Universit{\"a}t Bochum, Universit{\"a}tsstra{\ss}e 150, D-44801 Bochum, Germany \\
$^{5}$ Harvard-Smithsonian Center for Astrophysics, 60 Garden Street, Cambridge, MA 02138, USA \\
$^{6}$ CSIRO Astronomy \& Space Science, P.O. Box 76, Epping, NSW 1710, Australia \\
$^{7}$ School of Physics \& Astronomy, Cardiff University, Queens Buildings, The Parade, Cardiff CF24 3AA, UK \\
}

\pubyear{2016}

\begin{document}


\label{firstpage}
\pagerange{\pageref{firstpage}--\pageref{lastpage}}

\maketitle

\begin{abstract}

We examine the processes triggering star formation and Active Galactic Nucleus (AGN) activity in a sample of 25 low redshift ($z<0.13$) gas-rich galaxy mergers observed at milli-arcsecond resolution with Very Long Baseline Interferometry as part of the mJy Imaging VLBA Exploration at 20cm (mJIVE-20) survey. The high ($>10^7$\,K) brightness temperature required for an mJIVE-20 detection allows us to unambiguously identify the radio AGN in our sample. We find three such objects. Our VLBI AGN identifications are classified as Seyferts or LINERs in narrow line optical diagnostic plots; mid-infrared colours of our targets and the comparison of H$\alpha$ star formation rates with integrated radio luminosity are also consistent with the VLBI identifications. We reconstruct star formation histories in our galaxies using optical and UV photometry, and find that these radio AGN are not triggered promptly in the merger process, consistent with previous findings for non-VLBI samples of radio AGN. This delay can significantly limit the efficiency of feedback by radio AGN triggered in galaxy mergers. We find that radio AGN hosts have lower star formation rates than non-AGN radio-selected galaxies at the same starburst age. Conventional and VLBI radio imaging shows these AGN to be compact on arcsecond scales. Our modeling suggests that the actual sizes of AGN-inflated radio lobes may be much larger than this, but these are too faint to be detected in existing observations. Deep radio imaging is required to map out the true extent of the AGN, and to determine whether the low star formation rates in radio AGN hosts are a result of the special conditions required for radio jet triggering, or the effect of AGN feedback.


\end{abstract}


\section{Introduction}
\label{sec:intro}

The growth of galaxies and supermassive black holes at their centres are tightly coupled. Black hole masses are closely correlated with the stellar properties of their host galaxies (Magorrian et al. 1998, H\"aring \& Rix 2004, G\"ultekin et al. 2009), and the Active Galactic Nucleus (AGN) fraction and cosmic star formation rate density both peak around $z \sim 2-3$ (Hopkins \& Beacom 2006, Richards et al. 2006). Observationally, supermassive black holes are seen to impart thermal and kinetic feedback on their host galaxies and beyond. Silk \& Rees (1998) highlighted that the radiation pressure from an AGN fed at the Eddington limit can significantly limit accretion onto its host galaxy. The predicted scaling between black hole mass and stellar velocity dispersion in the AGN host depends on the details of the feedback process, including whether it is energy or momentum driven (King 2003) and the role played by dust grains (Fabian et al. 2008). AGN winds provide another form of feedback; wind kinetic luminosities of 5-10\% of the Eddington luminosity have been reported (Dunn et al. 2010, Tombesi et al. 2012). Large-scale AGN-driven outflows at both low (Sturm et al. 2011, Greene et al. 2012) and high (e.g. in a $z=6.4$ quasar; Maiolino et al. 2012) redshifts have been observed. These often shows mass outflow rates in excess of 1000\,$M_\odot$\,yr$^{-1}$ and velocities in excess of 1000\,km\,s$^{-1}$. Such outflows have also been reported in galaxies hosting powerful radio AGN at both low (Holt et al. 2008, Morganti et al. 2015) and high (Nesvadba et al. 2008) redshifts.


Number densities of luminous AGN decrease rapidly after the peak of AGN and star formation activity at $z \sim 2-2.5$ (e.g. Ueda et al. 2003); this is known as ``AGN downsizing'' (see Fabian 2012 for a review). On the other hand, large populations of radiatively inefficient AGN with synchrotron emission observable in the radio part of the electromagnetic spectrum emerge over the last half of the Hubble time (e.g. Best \& Heckman 2012; see Heckman \& Best 2014 for a review). Galaxy formation models routinely invoke so-called ``maintenance mode'' AGN feedback in order to explain the observed truncation of star formation in galaxies since $z \sim 1$ (Croton et al. 2006, Bower et al. 2006, Shabala \& Alexander 2009)\footnote{Although AGN feedback is usually invoked to suppress star formation, we note that this feedback can also be positive, for example as seen in the filaments of Centaurus A (Oosterloo \& Morganti 2005, Crockett et al. 2012) and in Minkowski's object (Croft et al. 2006), where the passage of the AGN jets and/or bow shocks {\it triggers} star formation.}. This feedback is supplied through radio jets impacting on the hot halo gas around massive elliptical galaxies, affecting both the host galaxy (Morganti et al. 2013) and larger scale environment (Fabian et al. 2003, Shabala et al. 2011, McNamara \& Nulsen 2012). Numerous studies (e.g. Rafferty et al. 2006, Best et al. 2007; see Cattaneo et al. 2009 for a review) have argued that the rate at which AGN jets supply energy approximately offsets cooling losses. The radio AGN activity is intermittent, as evidenced by observations of double-double radio sources, consisting of multiple pairs of jet-inflated radio lobes (Schoenmakers et al. 2000). 

AGN jet activity is powered by either accretion, black hole spin, or a combination of both processes (Meier 2001, Benson \& Babul 2009). Unlike their lower mass counterparts, massive galaxies -- which host the bulk of AGN in the local Universe -- are relatively poor in cold gas (Kannappan 2004), and the fuel required to power the AGN can come from gentle cooling of the hot gas in the halo surrounding the host galaxy; or dynamical processes such as galaxy interactions. Hardcastle et al. (2007) suggested that these two scenarios can be observationally distinguished as low (LERG) and high-excitation (HERG) radio galaxies respectively. Best \& Heckman (2012) confirmed this hypothesis by showing that the LERG and HERG populations inhabit different environments. In particular, high-excitation radio AGN are found in lower-mass hosts and in more isolated environments, consistent with the idea that these AGN are triggered by galaxy interactions rather than cooling of hot X-ray emitting gas.

Galaxy interactions (and minor mergers in particular) contribute significantly to both galaxy growth (Aumer et al. 2014, Kaviraj 2014) and the cosmic star formation rate (Kaviraj et al. 2013) since $z \sim 2$. Thus, on the one hand minor mergers trigger star formation; on the other, they are also responsible for AGN activity which can inhibit star formation. The exact chronology of the processes driving star formation and AGN activity in mergers is crucial: if the AGN switches on after the bulk of the merger-triggered star formation has already taken place, the efficiency of AGN feedback in truncating star formation in HERGs will be severely limited.

Previous studies have used emission-line diagnostics (Baldwin et al. 1981, Kauffmann et al. 2003, Kewley et al. 2006) to distinguish between AGN and star formation dominated systems. Proceeding in this way, Schawinski et al. (2007) and Wild et al. (2010) found that emission-line AGN have stellar populations $200-300$~Myr older than vigorously star-forming galaxies. Schawinski et al. considered 16,000 early-type galaxies at $0.005 < z< 0.1$; while Wild et al. examined only post-starburst galaxies in the Sloan Digital Sky Survey (SDSS; Abazajian et al. 2009). Shabala et al. (2012) examined the largest morphologically-selected sample of local galaxies with prominent dust lanes (Kaviraj et al. 2012), compiled as part of the Galaxy Zoo 2 project\footnote{This publication has been made possible by the participation of more than 250 000 volunteers in the Galaxy Zoo 2 project. Their contributions are individually acknowledged at http://zoo2.galaxyzoo.org/authors}. Over two thirds of the observed galaxies were found to be morphologically disturbed even in shallow (54s exposure) SDSS images (Kaviraj et al. 2012), and showed elevated (by a factor of 3) levels of both star formation and AGN activity compared to matched samples of similar galaxies without dust lanes (Shabala et al. 2012). These authors concluded that their dust lane galaxies were remnants of gas-rich minor mergers, and therefore are ideal laboratories for studying the co-evolution of star formation and AGN activity. By reconstructing photometric star formation histories for these galaxies, they also found that AGN hosts have stellar populations that are a few hundred Myr old, broadly consistent with the results of Schawinski et al. (2007) and Wild et al. (2010). A broader study of morphologically selected low-redshift mergers by Carpineti et al. (2012) found the related result that while the fraction of star-forming galaxies peaks in systems undergoing mergers, the optical AGN fraction peaks in post-merger remnants.

Importantly, the Shabala et al. (2012) AGN classification was based on radio continuum imaging rather than optical emission line luminosity, potentially probing a later phase in post-merger evolution of the host galaxy (Cowley et al. 2016). Unlike single-fibre optical spectroscopic diagnostics, radio continuum images also provide spatial information about the extent of the radio jets. Extended jets are indicative of older AGN, which have potentially imparted significant feedback on star formation in the host galaxy. Radio emission in galaxies can come about either due to AGN jets or supernova-driven shocks. In Shabala et al. (2012), AGN were identified as those galaxies in which the radio luminosity (drawn from the Faint Images of the Radio Sky at Twenty Centimetres, FIRST, survey; Becker et al. 1995) significantly (1.5$\sigma$) exceeded the SDSS star formation rate (Brinchmann et al. 2004). While useful, this separation can suffer from misclassification due to the scatter in the star formation rate -- radio luminosity relation. In particular, it is biased against galaxies with comparable levels of radio emission coming from star formation and the AGN. Moreover, the 5.4 arcsec resolution of FIRST is comparable with the sizes of a number of galaxies in the sample, and it is therefore unclear whether the AGN has imparted any feedback on smaller scales.

To address these concerns, we present high-resolution Very Long Baseline Inteferometry (VLBI) radio observations of the Shabala et al. (2012) dust lane galaxies. In Section~\ref{sec:VLBI_intro} we describe the observations, and present results in Section~\ref{sec:results}. We outline our VLBI analysis procedure in Section~\ref{sec:analysis}, compare optical, radio, infra-red and X-ray AGN diagnostics in Section~\ref{sec:AGN_comparison}, examine star formation rates in Section~\ref{sec:SFRs}, and reconstruct star formation histories for our galaxies in Section~\ref{sec:SFH}. We discuss the implications of our findings on the relationship between star formation and AGN triggering in Sections~\ref{sec:lobeLumins} and \ref{sec:effFeedback}, and conclude in Section~\ref{sec:conclusions}.

We use $H_0 = 67.8$\,km\,s$^{-1}$\,Mpc$^{-1}$ and $\Omega_{\rm M}=0.308$ (Plank Collaboration, 2016) throughout the paper.

\section{Widefield VLBI imaging of dust lane galaxies}
\label{sec:VLBI_intro}

Kaviraj et al. (2012) and Shabala et al. (2012) presented a sample of 484 dust lane galaxies at $0.01<z<0.15$. These were objects flagged by at least one Galaxy Zoo 2 user as containing a dust feature. Each galaxy in this sample was then visually re-inspected by two of us (S.K. and Y.S.T.) to determine whether the galaxy did indeed have a dust lane. For further details of this selection process we refer the reader to Kaviraj et al. (2012). For the purposes of this work, we note that 352 of the dust lane galaxies (73\%) were classified as early-type, with the rest being bulge-dominated Sa galaxies.

126 confirmed dust lane galaxies had FIRST 1.4 GHz radio counterparts within a 3 arcsec radius. These galaxies were selected for follow up with widefield VLBI imaging under the auspices of the mJIVE-20 project (Deller \& Middelberg 2014). The mJy Imaging VLBA Exploration at 20cm survey is using the VLBA to systematically observe objects detected by the FIRST survey. By utilising short segments scheduled in bad weather or with a reduced number of antennas, the mJIVE-20 survey has imaged more than 25,000 FIRST sources, with more than 5000 VLBI detections. The median sensitivity of mJIVE-20 is 1.2~mJy/beam, and median resolution is $6 \times 17$~milli-arcsecond, corresponding to a brightness temperature of $>10^7$~K; the typical field-to-field variation in these values is approximately a factor of two. The largest angular size to which the VLBA is sensitive at 1.4 GHz is $\sim 0.3$~arcsec, set by its shortest baseline between Los Alamos and Pie Town; this corresponds to 713~pc for the highest redshift in our sample. We refer readers to Deller \& Middelberg (2014) for further details of the survey.

Widefield VLBI imaging with mJIVE-20 relies on the capability of the DiFX correlator (Deller et al. 2007) to place multiple phase centres within a large fraction of the primary beam of a VLBA antenna. Specifically, phase centres were placed at locations of all FIRST sources within $\sim 20$~arcmin of the beam centre. By design, each mJIVE-20 pointing encompassed a known VLBA calibrator with a minimum 1.4 GHz flux density of 100~mJy, meaning continuous in-beam calibration was possible. 25 of the 126 dust lane galaxies with a FIRST counterpart were located sufficiently near a known calibrator, and all were observed with the VLBA using the standard setup (64 MHz bandwidth, dual polarisation, yielding a total data rate of 512 Mbps). Depending on the location of the target with respect to the calibrator, the target on-source time ranged from 13 to 52 minutes, with a 1$\sigma$ rms ranging from 0.07 to 0.36 mJy/beam. A 6.75$\sigma$ detection threshold is adopted for a VLBI detection; in other words, the FIRST target is considered to have no VLBI detection if none of the pixels in the 4096x4096 pixel VLBA image exceed 6.75$\sigma$. This value was shown by Deller \& Middelberg (2014) to optimise the completeness of mJIVE-20 while still maintaining a low rate (0.3\%) of false positives. Our observations are summarised in Table~\ref{tab:observations}.

\begin{table*}
\centering
\tiny
\tabcolsep=0.11cm
\begin{tabular}{ccccccccccccccc}
\hline
& FIRST &&&&& mJIVE &&&&&&& \\
\hline
Name			& $S_{\rm peak}$	&	$S_{\rm int}$	&	$\sigma$	&	$\theta_{\rm maj} \times \theta_{\rm min}$			& pos. angle		& mJIVE-20	&	$S_{\rm peak}$	&	$S_{\rm int}$	&	$\sigma$	&	VLBI & beam size					& Class 			& log $L_{\rm 1.4, vlbi}$	\\
			& (mJy/bm)	&	(mJy)	&	(mJy/bm)	&	(arcsec)			& (degrees)		& name	&	(mJy/bm)	&	(mJy) &		(mJy/bm)	&	fraction	 & (mas)					&			& (W/Hz)	\\
(1) & (2) & (3) & (4) & (5) & (6) & (7) & (8) & (9) & (10) & (11) & (12) & (13) & (14) \\ 																																
\hline																																
2MASX	J03004681	&	6.63	&	6.58	&	0.11	& $	1.4	\times	0	$ &	56	&	MJV02278	&	5.07	&	5.88	&	0.25	&	0.89	&	21.3	x	7.8	&	AGN	&	23.33	\\
-0001556 &&&&&&&&&&\\																																
2MASX	J13153624	&	6.24	&	7.68	&	0.13	& $	2.9	\times	2.3	$ &	160	&	MJV16230	&	2.6	&	2.81	&	0.21	&	0.37	&	15	x	6.9	&	AGN	&	22.44	\\
+4113271 &&&&&&&&&&\\																																
UGC	5498	&	22.04	&	23.34	&	0.15	& $	1.5	\times	1.1	$ &	114	&	MJV16427	&	8.3	&	15.65	&	0.16	&	0.67	&	15.1	x	6.5	&	AGN	&	22.18	\\
UGC	9639	&	17.88	&	30.41	&	0.14	& $	4.7	\times	4.3	$ &	23	&	MJV21402	&$<	0.74	$&	-	&	0.11	&$<	0.02	$&	18.8	x	5.6	&	SF dominant;	&$<	21.32	$\\
&&&&&&&&&&&&broadline AGN&\\
VCC	1802	&	7.64	&	11.37	&	0.16	& $	4.9	\times	2.6	$ &	63	&	MJV05787	&$<	0.58	$&	-	&	0.09	&$<	0.05	$&	16.3	x	6.6	&	SF dominant;	&$<	20.87	$\\
&&&&&&&&&&&&behind Virgo cluster&\\
2MASX	J16175244	&	6.2	&	6.85	&	0.14	& $	2.0	\times	1.4	$ &	100	&	MJV15923	&$<	0.57	$&	-	&	0.08	&$<	0.08	$&	16.1	x	6.5	&	SF dominant	&$<	21.26	$\\
+0604180 &&&&&&&&&&\\																																
2MFGC	9846	&	3.55	&	4.78	&	0.13	& $	4.8	\times	0.6	$ &	100	&	MJV21758	&$<	0.5	$&	-	&	0.07	&$<	0.10	$&	14.9	x	6.4	&	SF dominant	&$<	21.55	$\\
2MASX	J10462365	&	2.03	&	2.29	&	0.14	& $	4.0	\times	0	$ &	109	&	MJV09993	&$<	0.46	$&	-	&	0.07	&$<	0.20	$&	16.4	x	6.6	&	SF dominant	&$<	20.90	$\\
+0637104 &&&&&&&&&&\\																																
NGC	5145	&	4.56	&	29.12	&	0.14	& $	14.1	\times	11.1	$ &	92	&	MJV16966	&$<	1.15	$&	-	&	0.17	&$<	0.04	$&	18.2	x	6.8	&	SF dominant	&$<	19.61	$\\
UGC	7098	&	1.84	&	8.95	&	0.14	& $	13.6	\times	8.0	$ &	103	&	MJV21480	&$<	1.48	$&	-	&	0.22	&$<	0.17	$&	15.2	x	6.5	&	SF dominant	&$<	21.65	$\\

2MASX	J16284296	&	3.27	&	4.35	&	0.15	& $	4.5	\times	1.3	$ &	30	&	MJV16564	&$<	1.23	$&	-	&	0.18	&$<	0.28	$&	14.2	x	5.9	&	SF probable	&$<	21.49	$\\
+2223488 &&&&&&&&&&\\																																
UGC	10205	&	1.9	&	2.43	&	0.14	& $	4.6	\times	0	$ &	144	&	MJV16926	&$<	0.73	$&	-	&	0.11	&$<	0.30	$&	17.2	x	7	&	SF probable	&$<	20.89	$\\
2MASX	J09192731	&	2.53	&	3.63	&	0.14	& $	5.2	\times	1.3	$ &	165	&	MJV16796	&$<	1.03	$&	-	&	0.15	&$<	0.28	$&	15	x	6.3	&	SF probable	&$<	20.91	$\\
+3347270 &&&&&&&&&&\\																																
2MASX	J16075348	&	1.65	&	1.85	&	0.16	& $	2.3	\times	1.3	$ &	0	&	MJV01278	&$<	0.74	$&	-	&	0.11	&$<	0.40	$&	15.7	x	6.6	&	SF probable	&$<	21.25	$\\
+1016098 &&&&&&&&&&\\																																
2MASX	J13135648	&	1.38	&	3.13	&	0.14	& $	6.3	\times	5.9	$ &	93	&	MJV16360	&$<	0.92	$&	-	&	0.14	&$<	0.29	$&	19	x	6.2	&	SF probable	&$<	21.93	$\\
+5326512 &&&&&&&&&&\\																																
UGC	9014	&	1.01	&	2.56	&	0.15	& $	10.2	\times	3.5	$ &	70	&	MJV21867	&$<	1.08	$&	-	&	0.16	&$<	0.42	$&	15.1	x	6.2	&	SF probable	&$<	20.67	$\\
2MASX	J11433236	&	1.34	&	3.85	&	0.15	& $	12.7	\times	2.8	$ &	152	&	MJV21996	&$<	1.5	$&	-	&	0.22	&$<	0.39	$&	27.6	x	10.6	&	SF probable	&$<	21.97	$\\
+1541123 &&&&&&&&&&\\																																
FGC	1015	&	1.03	&	1.23	&	0.14	& $	10.0	\times	2.6	$ &	20	&	MJV09889	&$<	0.82	$&	-	&	0.12	&$<	0.67	$&	16.8	x	7.1	&	SF plausible	&$<	21.24	$\\
MCG	-116	&	1.11	&	1.43	&	0.17	& $	7.8	\times	0	$ &	156	&	MJV00640	&$<	0.96	$&	-	&	0.14	&$<	0.67	$&	16.3	x	6	&	SF plausible	&$<	21.31	$\\
																																
2MASX	J10272554	&	1.48	&	3.67	&	0.15	& $	9.3	\times	4.0	$ &	32	&	MJV22110	&$<	2.07	$&	-	&	0.31	&$<	0.56	$&	15.3	x	6.4	&	SF plausible	&$<	22.26	$\\
+4735490 &&&&&&&&&&\\																																
2MASX	J08120662	&	2.33	&	1.67	&	0.19	& $	0	\times	0	$ &	5	&	MJV16644	&$<	0.67	$&	-	&	0.1	&$<	0.40	$&	16.5	x	5.9	&	unknown	&$<	21.44	$\\
+5713008 &&&&&&&&&&\\																																
CGCG	288-011	&	1.66	&	1.71	&	0.18	& $	3.8	\times	0	$ &	151	&	MJV13937	&$<	0.88	$&	-	&	0.13	&$<	0.51	$&	17.7	x	5.7	&	unknown	&$<	20.58	$\\
CGCG	270-035	&	1.71	&	1.14	&	0.13	& $	0	\times	0	$ &	103	&	MJV21693	&$<	1.1	$&	-	&	0.16	&$<	0.96	$&	16.1	x	6	&	unknown	&$<	21.45	$\\
2MASX	J11230330	&	1.34	&	1.45	&	0.15	& $	3.7	\times	0	$ &	47	&	MJV16733	&$<	2.4	$&	-	&	0.36	&$<	1.66	$&	16.4	x	7.5	&	unknown	&$<	22.44	$\\
+5957112 &&&&&&&&&&\\																																
IC	1182	&	1.62	&	2.51	&	0.15	& $	5.0	\times	3.0	$ &	130	&	MJV21592	&$<	1.49	$&	-	&	0.22	&$<	0.59	$&	18.2	x	5.7	&	unknown;	&$<	21.58	$\\
&&&&&&&&&&&&X-ray quasar&\\
\hline
\end{tabular}
\caption{Dust lane galaxies observed with the VLBA. (1) Galaxy name. (2) 1.4 GHz FIRST peak flux density (Becker et al. 1995). (3) FIRST integrated flux density. (4) FIRST image rms. (5) deconvolved major and minor FIRST axes. (6) Major axis position angle. (7) mJIVE-20 name. (8) 1.4 GHz VLBI peak brightness. (9) VLBI total flux density. (10) VLBI image rms. (11) fraction of FIRST integrated flux contained in VLBI integrated flux, i.e. column 9 divided by column 3 for VLBI detections, and column 8 divided by column 3 for non-detections. (12) VLBI synthesised beam size. (13) Classification. (14) 1.4 GHz VLBI luminosity.}
\label{tab:observations}
\end{table*}


We note that, by only selecting galaxies with detected FIRST emission, we are by construction biased towards non-quiescent galaxies; in other words, objects where one or both of AGN activity and star formation are present. However, there is no bias with respect to the targets' VLBI properties. Sources were scheduled for observation with the VLBA if a suitable in-beam calibrator could be found; this requirement amounts to a chance approximate line-of-sight alignment with a distant quasar. Importantly, presence of a VLBI core does not select against extended radio emission -- most low redshift ($z<0.3$) giant radio galaxies in the 3CRR survey have detected VLBI cores (Hardcastle et al. 1998). Hence, our VLBI-targeted subsample can be considered representative of radio-loud, gas-rich minor mergers in the local Universe. We note that none of our VLBI targets show extended radio FIRST emission on arcsecond scales which is not aligned with the optical galactic disk (see Table~\ref{tab:observations}), suggesting that no kpc-scale AGN jets are present; we return to this point in Section~\ref{sec:lobeLumins}.

\begin{figure*}
\begin{center}
\begin{minipage}{0.95\textwidth}
\includegraphics[width=0.23\textwidth]{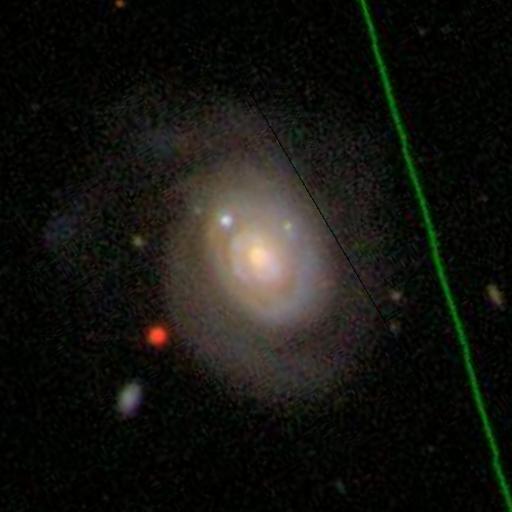}
\includegraphics[width=0.23\textwidth]{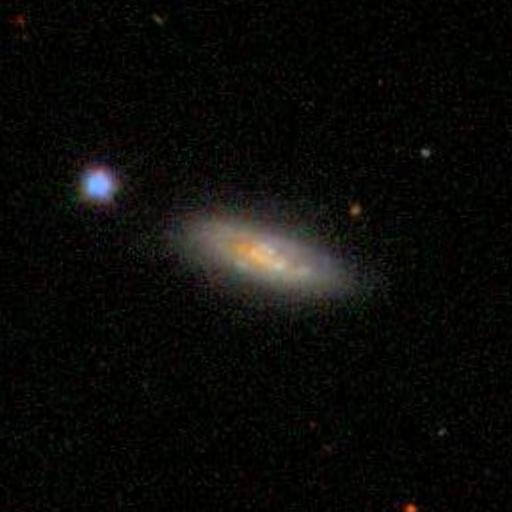}
\includegraphics[width=0.23\textwidth]{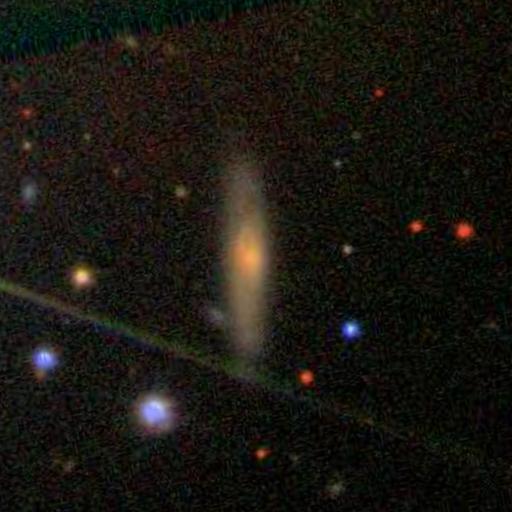}
\includegraphics[width=0.23\textwidth]{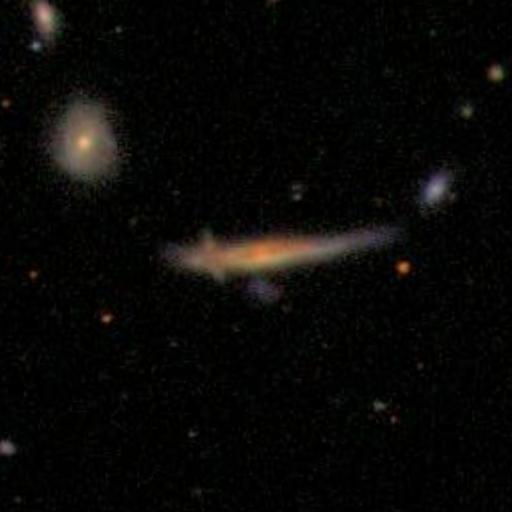}
\end{minipage}
\begin{minipage}{0.95\textwidth}
\includegraphics[width=0.23\textwidth]{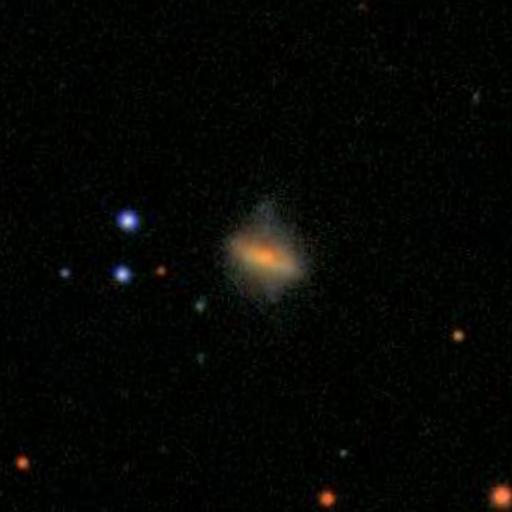}
\includegraphics[width=0.23\textwidth]{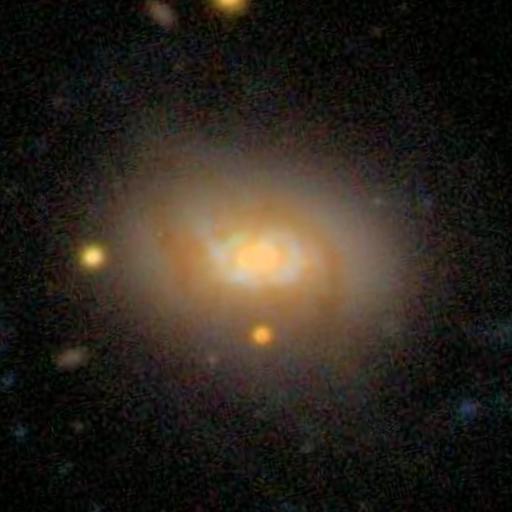}
\includegraphics[width=0.23\textwidth]{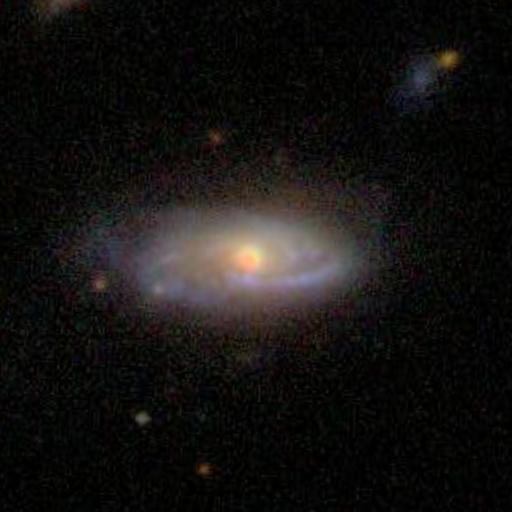}
\includegraphics[width=0.23\textwidth]{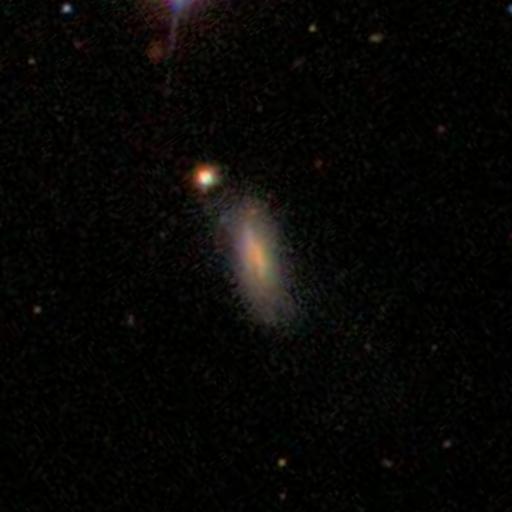}
\end{minipage}
\begin{minipage}{0.95\textwidth}
\includegraphics[width=0.23\textwidth]{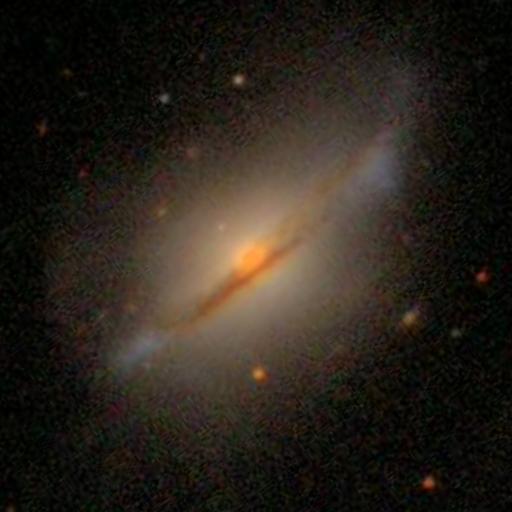}
\includegraphics[width=0.23\textwidth]{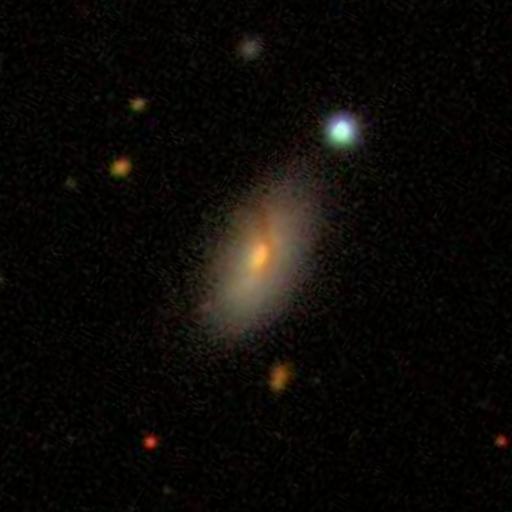}
\includegraphics[width=0.23\textwidth]{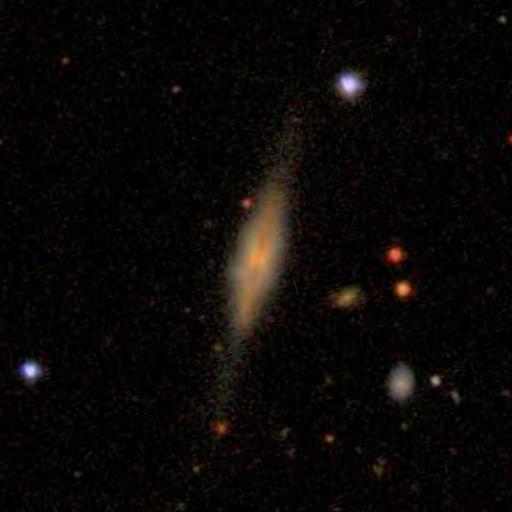}
\includegraphics[width=0.23\textwidth]{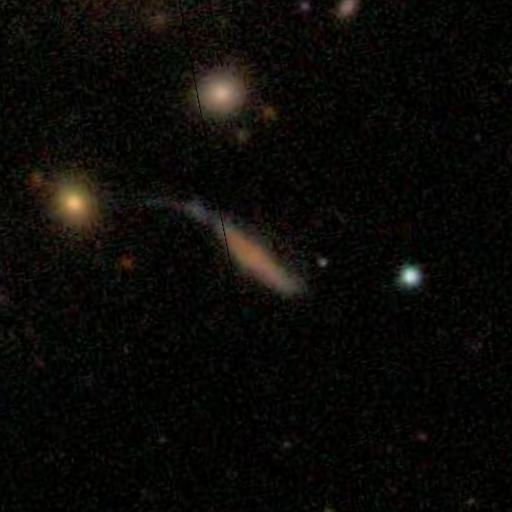}
\end{minipage}
\begin{minipage}{0.95\textwidth}
\includegraphics[width=0.23\textwidth]{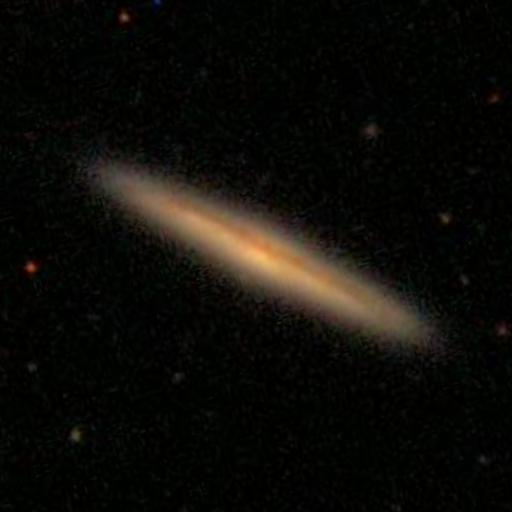}
\includegraphics[width=0.23\textwidth]{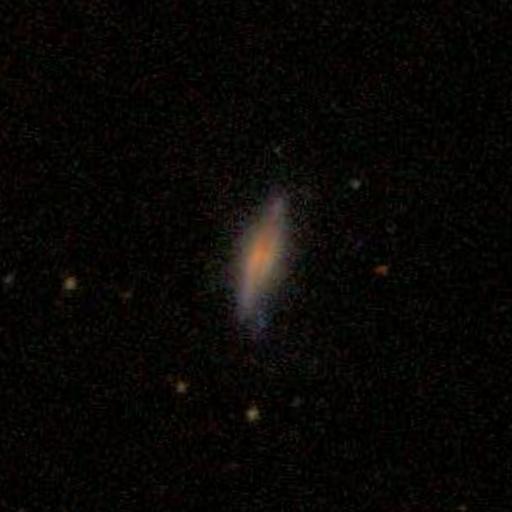}
\includegraphics[width=0.23\textwidth]{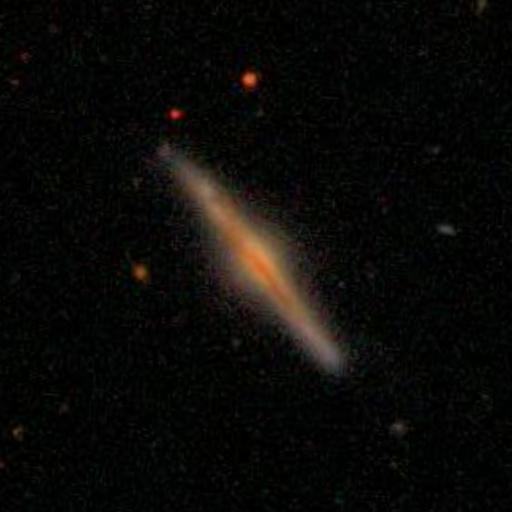}
\includegraphics[width=0.23\textwidth]{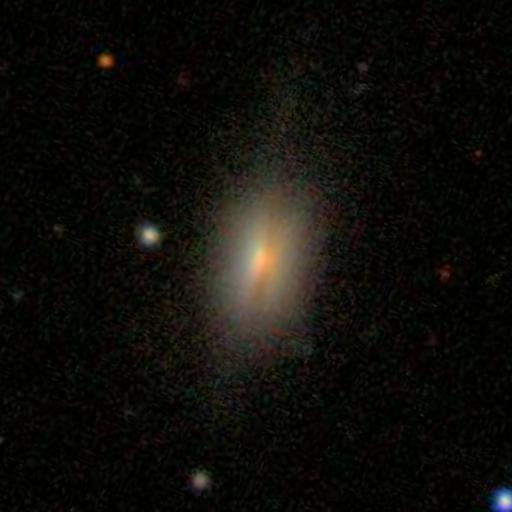}
\end{minipage}
\begin{minipage}{0.95\textwidth}
\includegraphics[width=0.23\textwidth]{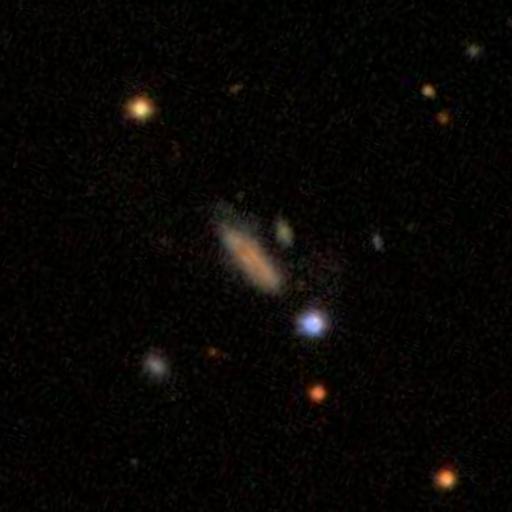}
\includegraphics[width=0.23\textwidth]{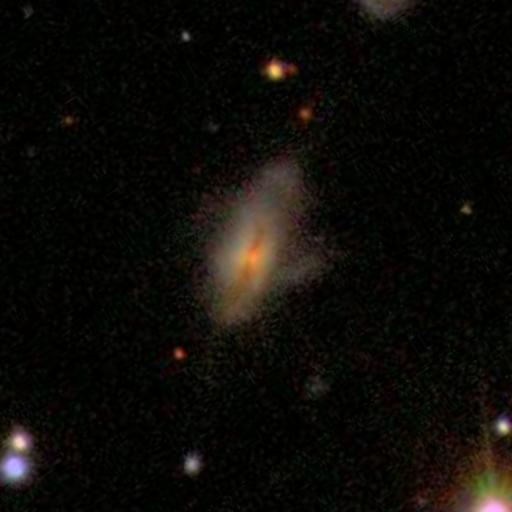}
\includegraphics[width=0.23\textwidth]{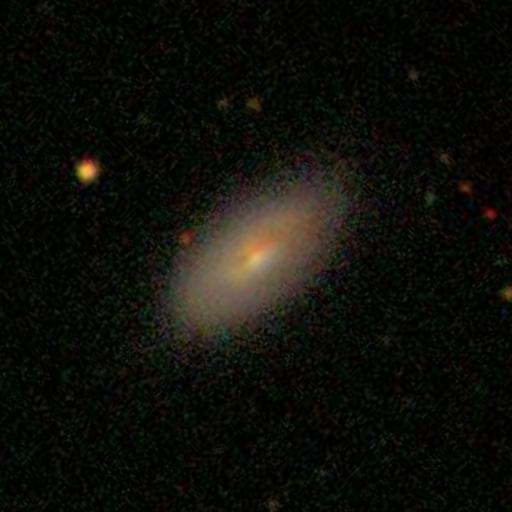}
\includegraphics[width=0.23\textwidth]{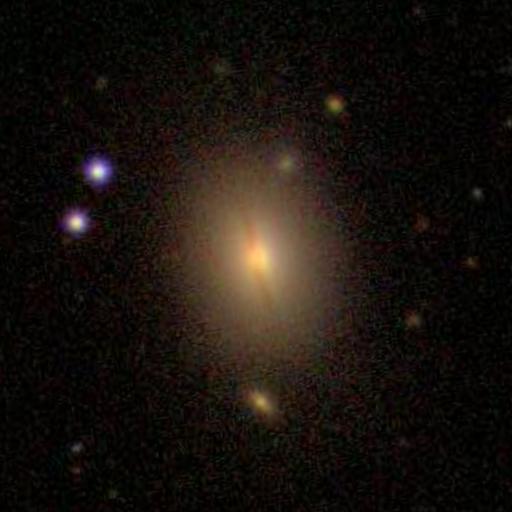}
\end{minipage}
\begin{minipage}{0.95\textwidth}
\includegraphics[width=0.23\textwidth]{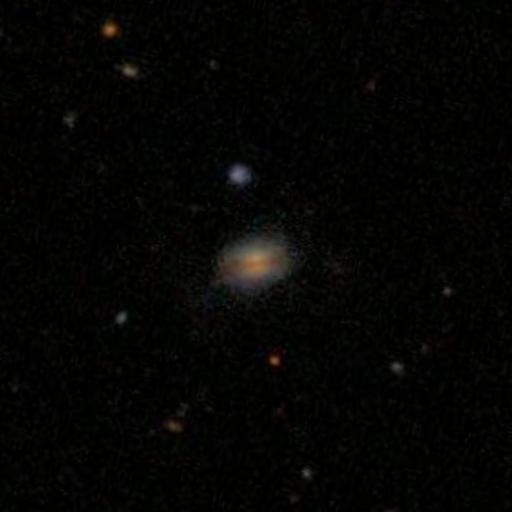}
\includegraphics[width=0.23\textwidth]{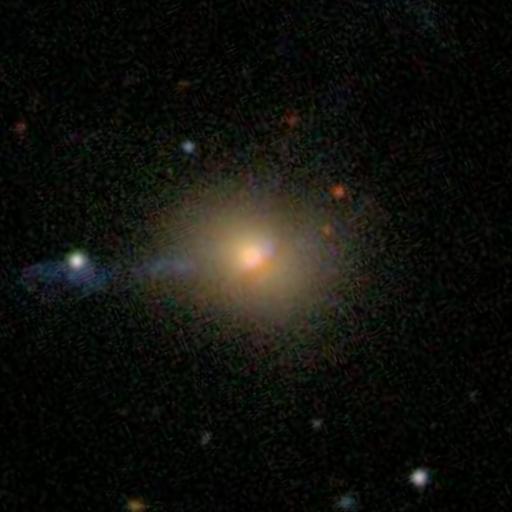}
\end{minipage}
\end{center}
\caption{Optical SDSS images of the 22 mJIVE-20 non-detections. The images are 100$\times$100 arcsec across. The order of images (starting top left, moving across, then down) is as in Table~\ref{tab:observations}.}
\label{fig:SDSS_sf}
\end{figure*}

\section{Results}
\label{sec:results}

\begin{figure*}
\includegraphics[width=0.3\textwidth,clip]{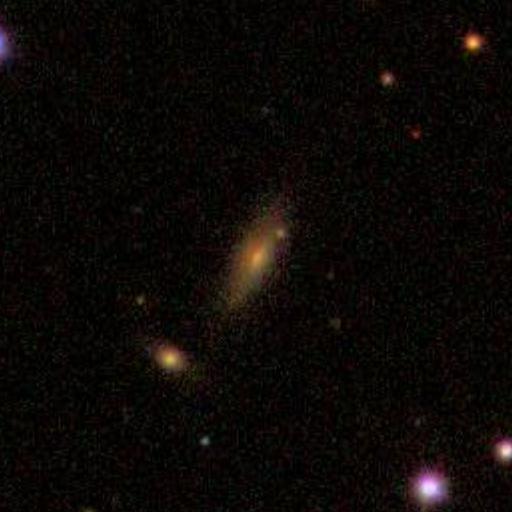}
\includegraphics[width=0.3\textwidth,clip]{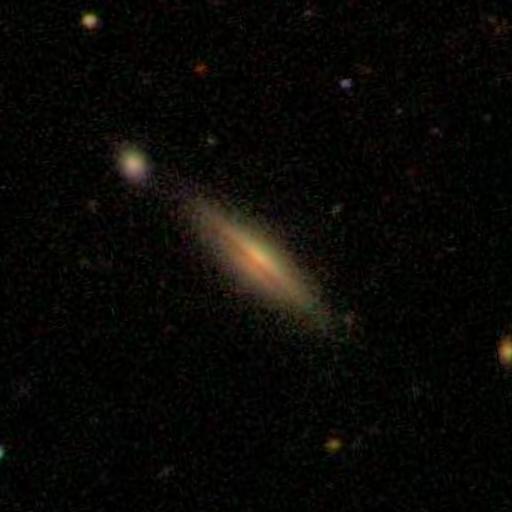}
\includegraphics[width=0.3\textwidth,clip]{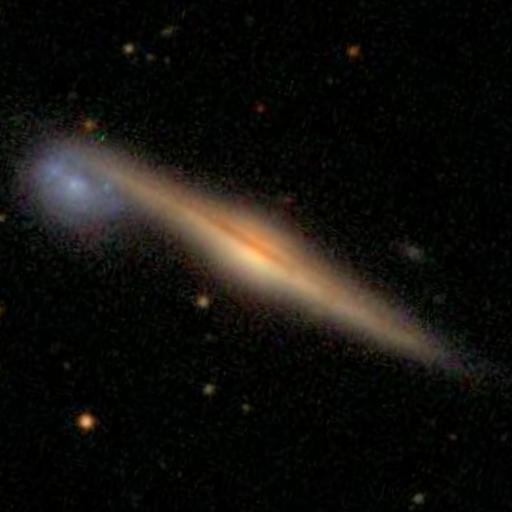}
\includegraphics[width=0.3\textwidth,clip,angle=0]{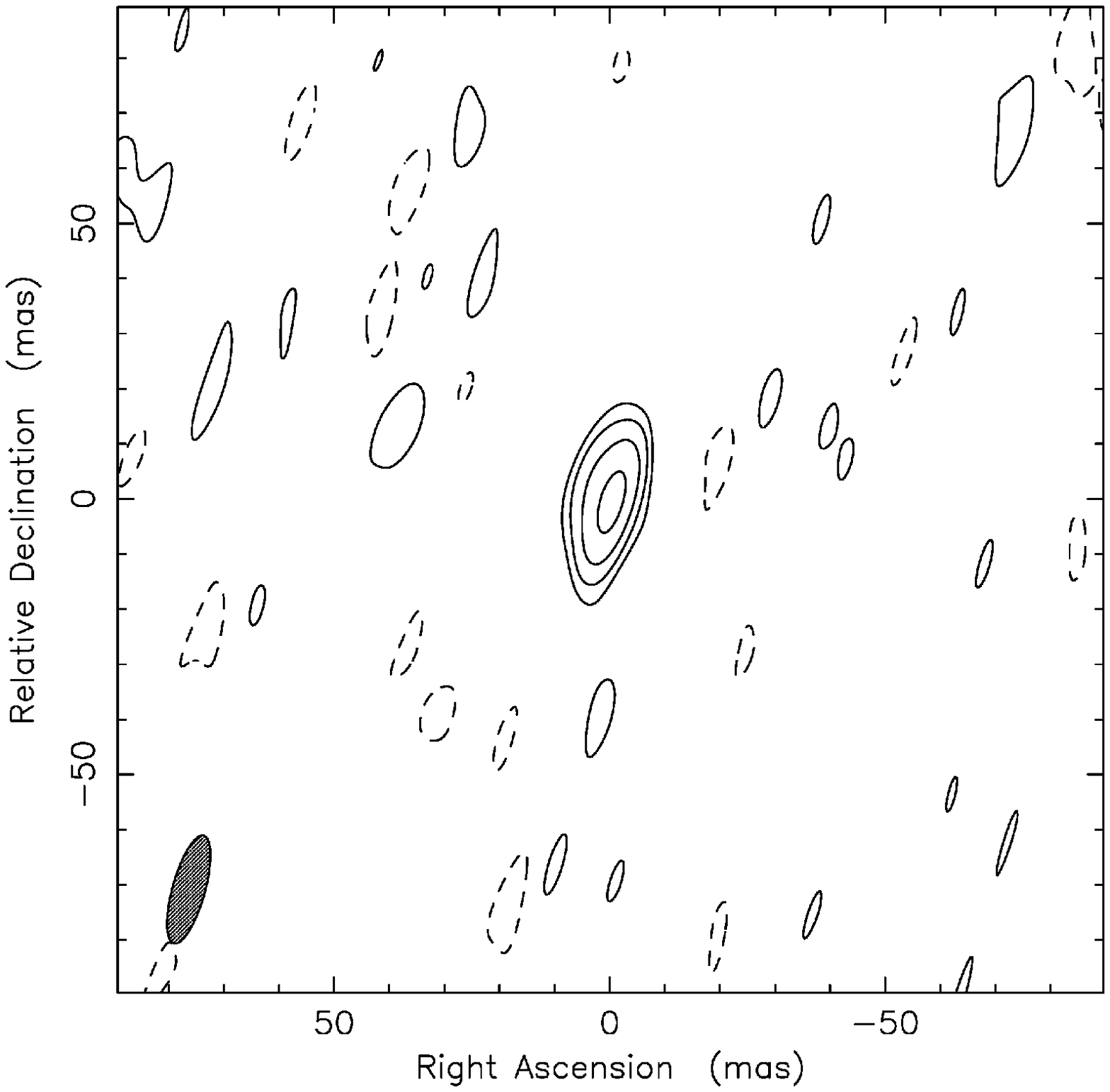}
\includegraphics[width=0.3\textwidth,clip,angle=0]{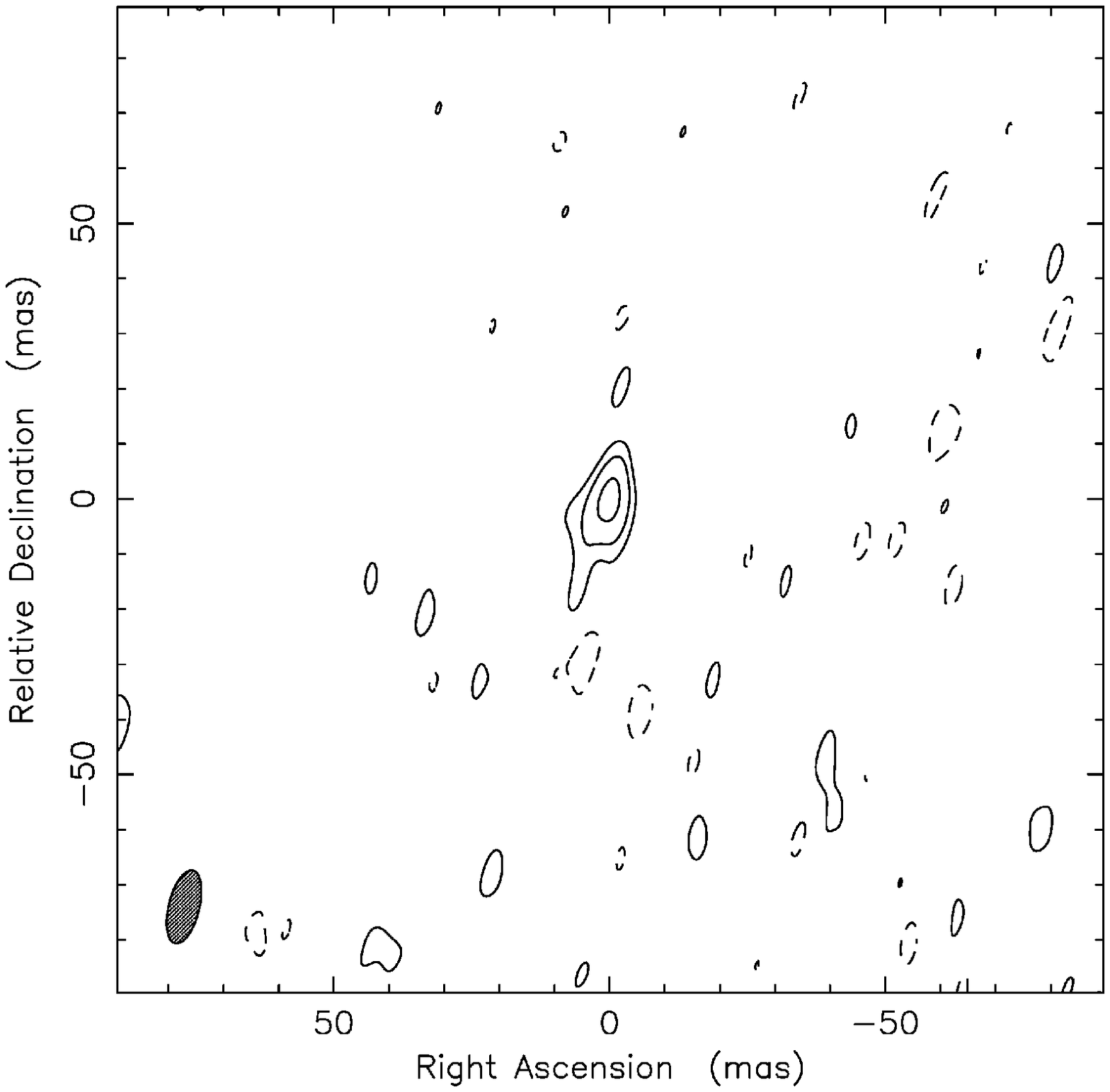}
\includegraphics[width=0.3\textwidth,clip,angle=0]{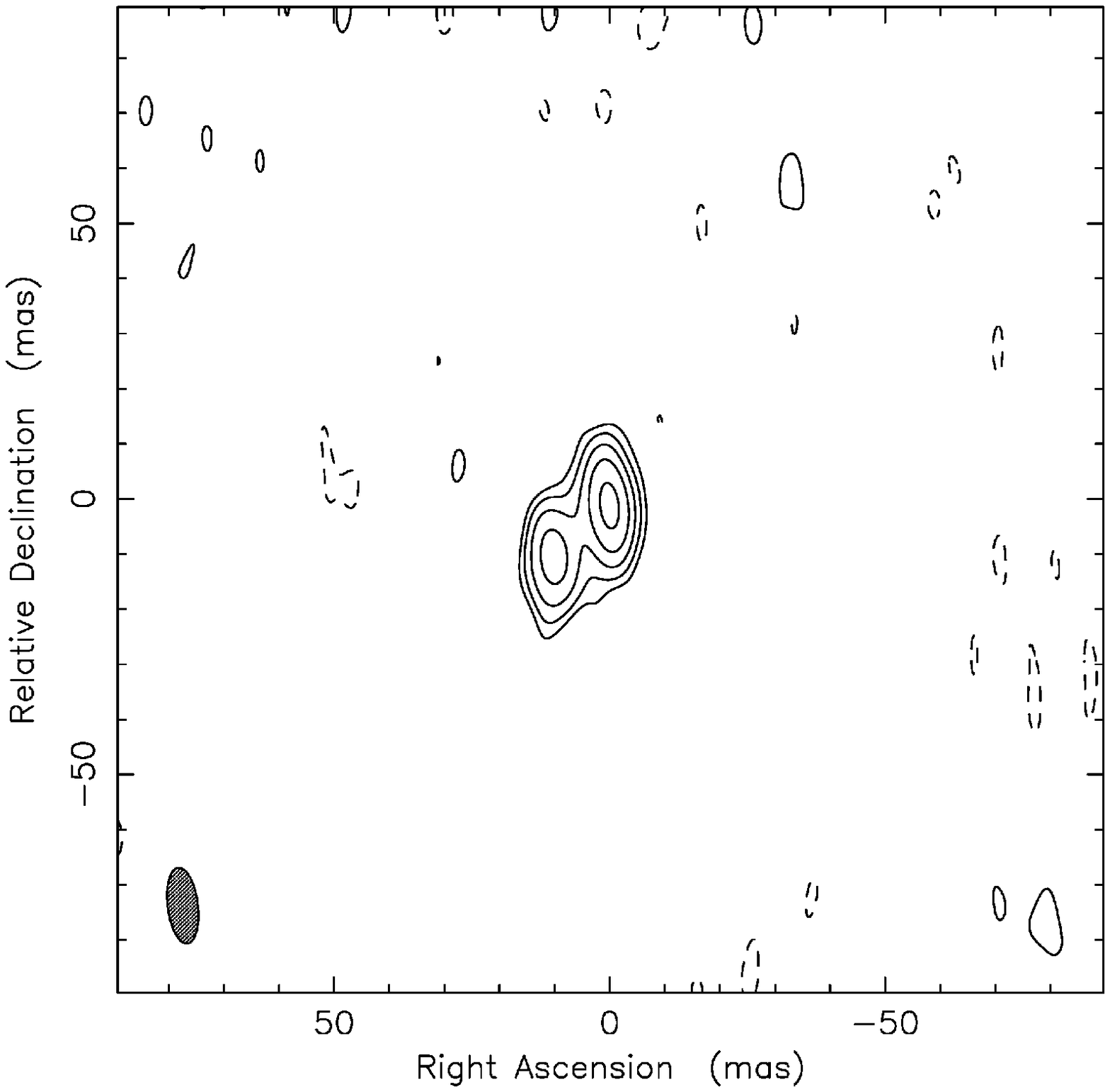}
\caption{Optical SDSS ({\it top}) and radio VLBI ({\it bottom}) images of the three dust lane mJIVE-20 detections. The optical images are 100$\times$100 arcsec across. The VLBI images are 0.2$\times$0.2 arcsec in size, i.e. 500 times smaller. {\it Left}: 2MASX\,J03004681-0001556; contours are at 10, 20, 40 and 80 percent of peak emission of 4.95~mJy/beam; {\it middle}: 2MASX\,J13153624+411327; contours are at 20, 40 and 80 percent of peak emission of 2.51~mJy/beam; {\it right}: UGC\,05498; contours are at 5, 10, 20, 40 and 80 percent of peak emission of 8.39~mJy/beam. In the case of UGC\,05498, a clear compact double radio source is visible, oriented perpendicular to the galactic disk.}
\label{fig:AGN}
\end{figure*}

The results of our VLBI imaging are shown in Table~\ref{tab:observations}. Table~\ref{tab:observed_sources} presents the optical properties of our target galaxies, and their optical morphologies are shown in Figures~\ref{fig:SDSS_sf} and \ref{fig:AGN}.
Three galaxies were detected to have high-resolution VLBI cores, strongly suggesting these are AGN\footnote{The brightest supernova or supernova remnants in the prototypical ULIRG Arp220, located at $z=0.018$, have VLBI peak flux  densities of $\leq 1$~mJy/beam (Batejat et al. 2011). At this redshift, our two compact VLBI detections would have peak flux densities of 9.7 and 36 mJy/beam, and therefore we rule them out as nuclear starburst activity.}. These are shown in Figure~\ref{fig:AGN}. One of the detections (UGC~05498) exhibits significant extended emission on VLBI scales, with a clear double-component structure oriented perpendicular to the optical major axis of the galaxy. The other two detections, 2MASX\,J03004681-0001556 and 2MASX\,J13153624+4113271, appear compact. 

The remaining 22 sources did not show any high brightness temperature components, placing upper limits on any AGN-related emission (column (8) in Table~\ref{tab:observations}). In some cases these upper limits are quite poor, sometimes even exceeding the FIRST peak flux density (see column (11) of Table~\ref{tab:observations}).

\section{Analysis}
\label{sec:analysis}

\subsection{VLBI detections}
\label{ref:analysis_detections}

For our VLBI detections, we used the {\texttt {blobcat}} package (Hales et al. 2012) to estimate source flux densities. {\texttt {Blobcat}} uses a flood-fill algorithm that selects a contiguous region of pixels, beginning with pixels above the $6.75\sigma$ threshold described in Section~\ref{sec:VLBI_intro}, and filling down to a second $3\sigma$ threshold. {\texttt {Blobcat}} then fits one or more Gaussian components to the selected pixels. To check our results, we also performed visibility model fitting using the {\texttt {difmap}} package (Shepherd 1997) for each of the three detections, using a single Gaussian component for 2MASX\,J03004681-0001556 and 2MASX\,J13153624+4113271, and two Gaussians plus a point source for UGC~05498; we recovered peak brightnesses and total flux densities that match the deconvolved image plane results to better than 10\% in all cases.

UGC~05498 clearly exhibits two components on VLBI scales, with physical separation of $\sim 10$\,pc. If these are not significantly Doppler boosted (i.e. the jets lie close to orthogonal to the line of sight), even under the conservative assumption that the radio jets expand at approximately the sound speed in the molecular gas ($\sim 10$ km/s), the implied age of these jets is $<10^6$ years. An alternative interpretation for the observed radio emission is as a Doppler-boosted one-sided jet with a compact core. Multi-frequency data is required to distinguish between these two scenarios: self-absorbed synchrotron cores of Doppler-boosted jets are expected to have both flat spectra and frequency-dependent positions (e.g. Lobanov 1998, Pushkarev et al. 2012), while genuine lobes should have steep spectra and frequency-invariant positions. We note that the observed radio structure is oriented perpendicular to the galactic stellar disk, consistent with other observations of VLBI-scale jets in low redshift, massive late-type galaxies (Kaviraj et al. 2015b), and we therefore favour the non-beamed jet interpretation. 

The other two VLBI detections, 2MASX\,J03004681-0001556 and 2MASX\,J13153624+4113271 are both found at significantly higher redshifts. 2MASX~J03004681-0001556 has 86 percent of its VLBI flux in the compact core, while 2MASX\,J13153624+4113271 contains 92 percent of its VLBI flux density in the core. Based on visibility model fitting, the sizes of these Gaussian components range from 4 to 6 milliarcseconds. It is not possible to estimate errors on the fitted sizes from a {\texttt difmap} model fit, however image plane deconvolution yields comparable best-fit sizes of $4-6$ mas, with uncertainties ranging from 0 milliarcseconds (i.e. we cannot rule out that the sources are unresolved) to approximately 8 milliarcseconds.

Assuming an upper limit of 8~milliarcseconds for the size of the VLBI cores, this corresponds to physical sizes of 19 and 11 pc for 2MASX\,J03004681-0001556 and 2MASX\,J13153624+4113271, respectively; we interpret these as likely upper limits on the sizes of the radio jets. We note that 2MASX\,J13153624+4113271 contains a significant amount of compact FIRST radio emission (55 percent) that is not captured in the mJIVE total flux. It is therefore possible that the true extent of the radio jets lies between the milliarcsecond scale morphology sampled by mJIVE and the arcsecond-scale resolution of FIRST. Alternatively, this emission could come from nuclear star formation in the host galaxy. 2MASX\,J13153624+4113271 would be an excellent target for an intermediate resolution interferometer such as eMERLIN. For the other two detections, total VLBI flux density makes up more than 70 percent of the compact FIRST flux density, suggesting that most of the observed arcsecond-scale flux is in fact found on milli-arcsecond scales.

\subsection{VLBI non-detections}
\label{ref:analysis_nonDetections}

The purpose of the VLBI observations is to identify AGN with luminosities comparable to any co-existing star formation in the host galaxy. We use the ratio of the upper limit on VLBI peak brightness and FIRST integrated flux densities, $S_{\rm VLBI, peak, upper} / S_{\rm FIRST, int}$, together with arcsecond-scale radio morphology from FIRST, to separate the non-detections into four groups as follows.

{\it SF dominant}: Objects in which the upper limit on VLBI integrated flux is $<0.25$ of the FIRST integrated flux are classified as ``star formation dominant''. We note that all three VLBI detections have $S_{\rm VLBI, int} / S_{\rm FIRST, int}>0.36$ and so fulfil this criterion comfortably. Five targets are classified as ``SF dominant''.

{\it SF probable}: Sources with $S_{\rm VLBI, peak, upper} / S_{\rm FIRST, int}$ in the range $0.25 - 0.5$ are classified as ``star formation probable''. We compared their deconvolved FIRST major and minor axes (Table~\ref{tab:observations}) to the optical SDSS images. Six of seven sources classified as ``SF probable'' showed resolved radio and optical structures which are aligned to better than 20 degrees (see Table~\ref{tab:observations} and Figure~\ref{fig:SDSS_sf}). The remaining target 2MASX\,J13135648+5326512 appears compact in both radio and optical images; it is classified as ``SF probable'' due to the low fraction of total flux density contained in the VLBI component ($S_{\rm VLBI, peak, upper} / S_{\rm FIRST, int}<0.29$) and consistency between radio and optical morphologies.

{\it SF plausible}: Three targets had poor VLBI upper limits, $S_{\rm VLBI, peak, upper} / S_{\rm FIRST, int} > 0.5$, while exhibiting resolved radio structures in FIRST. In all three cases, the radio major axis was aligned with the optical major axis of the host galaxy, indicating that the radio emission is consistent with originating from star formation. We classified these galaxies as ``star formation plausible''.

{\it Unknown}: Four targets were unresolved in FIRST. Three of these had poor upper limits on VLBI flux fraction, $S_{\rm VLBI, peak, upper} / S_{\rm FIRST, int} > 0.5$, and were classified as ``unknown''. The remaining target, 2MASX\,J08120662+5713008, has $S_{\rm VLBI, peak, upper} / S_{\rm FIRST, int} = 0.4$ but is resolved in the optical image; because of this mismatch between optical and radio morphologies we also classify this source as ``unknown''.

The remaining three targets were special cases. UGC\,9639, also known as Mrk\,834, is a well-known broad-line AGN. Contamination from the UV continuum of a Type 1 AGN may bias our stellar population estimates (Section~\ref{sec:SFH}), and we therefore exclude this object from analysis. VCC\,1802 is located behind the Virgo cluster. Potential contamination to its UV flux from nearby Virgo cluster galaxies seen in projection may again corrupt our star formation history estimates, and we therefore also exclude this object from further analysis. IC\,1182 (Mrk\,298; VLBI classification ``unknown'') is classified as a radio-quiet quasar by V\'eron-Cetty \& V\'eron (2006); however other studies have argued that this is also a starburst galaxy (Moles et al. 2004, Radovich et al. 2005). We retain it in our sample.

Below we compare a number of multi-wavelength AGN and star formation diagnostics, and conclude that generally non-detections classified as ``SF dominant'', ``SF probable'' or ``SF plausible'' are consistent with a star formation origin for the radio emission in these galaxies. The upper limits on 1.4\,GHz AGN luminosities set by our VLBI observations are $<10^{22}$\,W\,Hz$^{-1}$ for all classifications in these categories, with the exception of ``SF plausible'' galaxy 2MASX\,J10272554+4735490 (Table~\ref{tab:observations}).

\begin{figure}
\includegraphics[width=0.35\textwidth,clip,angle=270]{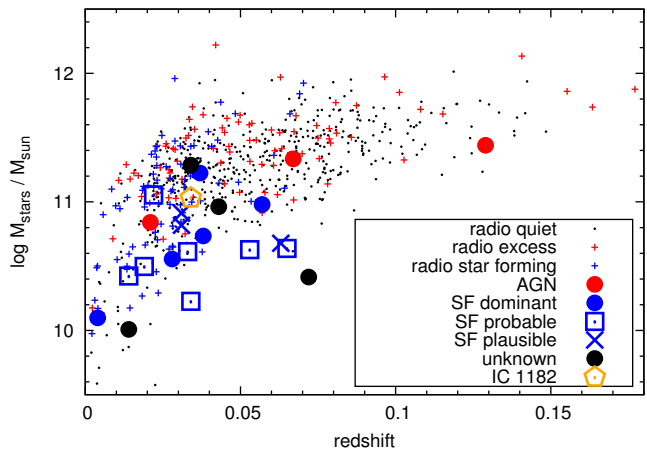}
\caption{Stellar masses and redshifts, obtained from the MPA-JHU catalogue, for the 23 classifiable galaxies observed in mJIVE-20. Both AGN and star-forming galaxies identified in mJIVE-20 (large symbols) follow similar a stellar mass - redshift relation to the parent population of Shabala et al. (2012) sample (small symbols). There appears to be no systematic offset in host properties of radio-quiet (black dots), radio excess (red crosses) and radio star forming (blue crosses) galaxies. Radio emission from star-forming galaxies can be detected reliably to lower redshifts than radio AGN emission.}
\label{fig:mass_redshift}
\end{figure}

In Figure~\ref{fig:mass_redshift} we plot the redshifts and stellar masses of our detections and non-detections. With the exception of 2MASX\,J03004681-0001556, a distant ($z=0.129$) and therefore massive galaxy, there appears to be no difference in the distributions of these properties for objects detected and not detected by mJIVE-20.

\begin{table}
\centering
\tiny
\tabcolsep=0.11cm
\begin{tabular}{ccccc}
\hline
Name		&	$z$	&	log SFR$_{\rm avg}$		&	log M$_\star$	& log $t_{\rm sb}$	\\
			&		&	(M$_{\odot}$/yr)		&	(M$_\odot$)	&	(Gyr)			\\
(1) & (2) & (3) & (4) & (5)\\ 
\hline
2MASX	J03004681	&	0.129	&	-0.37	&	11.44	& $	-0.2	^{+	0.2	}_{-	0.2	}$	\\
-0001556 &&&&\\															
2MASX	J13153624	&	0.067	&	-0.31	&	11.33	& $>	-0.17	^{+	0.23	}_{-	0.22	}$	\\
+4113271 &&&&\\															
UGC	5498	&	0.021	&	-0.24	&	10.84	& $>	-0.03	^{+	0.24	}_{-	0.27	}$	\\
2MASX	J16175244	&	0.038	&	0.46	&	10.73	& $	-0.56	^{+	0.23	}_{-	0.45	}$	\\
+0604180 &&&&\\															
2MFGC	9846	&	0.057	&	0.37	&	10.97	& $>	-0.38	^{+	0.08	}_{-	0.07	}$	\\
2MASX	J10462365	&	0.028	&	0.53	&	10.55	& $	-0.98	^{+	0.06	}_{-	0.05	}$	\\
+0637104 &&&&\\															
NGC	5145	&	0.004	&	-0.06	&	10.1	& $	-1.19	^{+	0.08	}_{-	0.07	}$	\\
UGC	7098	&	0.037	&	0.63	&	11.22	& $	-0.88	^{+	0.03	}_{-	0.04	}$	\\													
2MASX	J16284296	&	0.034	&	0.17	&	10.23	& $	-0.9	^{+	0.08	}_{-	0.1	}$	\\
+2223488 &&&&\\															
UGC	10205	&	0.022	&	0.19	&	11.05	& $	-0.15	^{+	0.14	}_{-	0.08	}$	\\
2MASX	J09192731	&	0.019	&	0.42	&	10.5	& $	-0.11	^{+	0.06	}_{-	0.07	}$	\\
+3347270 &&&&\\															
2MASX	J16075348	&	0.033	&	0.04	&	10.61	& $	0.01	^{+	0.22	}_{-	0.44	}$	\\
+1016098 &&&&\\															
2MASX	J13135648	&	0.065	&	0.12	&	10.64	& $	-0.61	^{+	0.04	}_{-	0.04	}$	\\
+5326512 &&&&\\															
UGC	9014	&	0.014	&	-0.86	&	10.42	& $	0.39	^{+	0.22	}_{-	0.42	}$	\\
2MASX	J11433236	&	0.053	&	1.06	&	10.63	& $	-0.73	^{+	0.11	}_{-	0.07	}$	\\
+1541123 &&&&\\															
FGC	1015	&	0.031	&	-0.19	&	10.91	& $	-0.89	^{+	0.1	}_{-	0.1	}$	\\
MCG	-116	&	0.031	&	-0.7	&	10.82	& $	0.23	^{+	0.24	}_{-	0.44	}$	\\															
2MASX	J10272554	&	0.063	&	0.51	&	10.68	& $	-0.6	^{+	0.05	}_{-	0.04	}$	\\
+4735490 &&&&\\															
2MASX	J08120662	&	0.043	&	0.77	&	10.96	& $>	-0.43	^{+	0.18	}_{-	0.12	}$	\\
+5713008 &&&&\\															
CGCG	288-011	&	0.014	&	-0.49	&	10.01	& $	-0.4	^{+	0.24	}_{-	0.45	}$	\\
CGCG	270-035	&	0.034	&	-0.33	&	11.29	& $	-0.31	^{+	0.23	}_{-	0.4	}$	\\
2MASX	J11230330	&	0.072	&	0.62	&	10.42	& $	-0.19	^{+	0.05	}_{-	0.05	}$	\\
+5957112 &&&&\\															
IC	1182	&	0.034	&	0.68	&	11.03	& $>	-0.34	^{+	0.37	}_{-	0.41	}$	\\													
\hline
\end{tabular}
\caption{Optical properties of VLBI targets. (1) Target name. (2) SDSS spectroscopic redshift (Salim et al. 2007). (3) Total SDSS star formation rate. (4) SDSS stellar mass. (5) Photometric starburst age.}
\label{tab:observed_sources}
\end{table}

\section{AGN diagnostics}
\label{sec:AGN_comparison}

\begin{figure}
\includegraphics[width=0.35\textwidth,clip,angle=270]{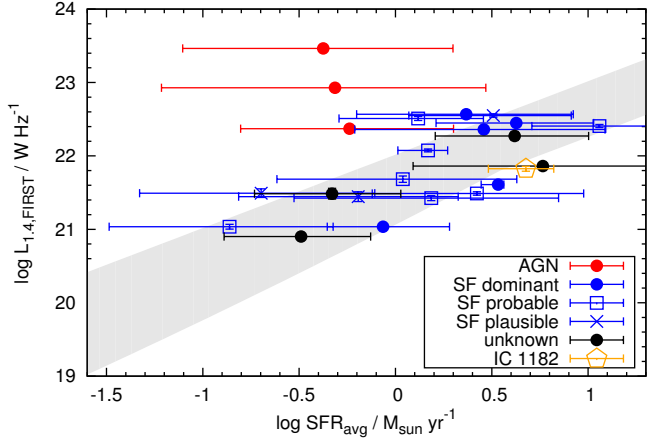}
\caption{1.4 GHz radio luminosity for the mJIVE-20 targets. Symbols are as in Figure~\ref{fig:mass_redshift}. Star formation rates are average values from the MPA-JHU catalogue (Brinchmann et al. 2004). The shaded region represents typical low-redshift SFR - $L_{1.4}$ relations, with the lower curve from Hopkins et al.'s (2003) fit to SDSS SFRs and the upper curve from Yun et al.'s (2001) results using IRAS data. VLBI detections lie above this curve.}
\label{fig:Lradio_excess}
\end{figure}

VLBI observations are by no means the only way of identifying whether our dust lane galaxies host AGN. In Shabala et al. (2012) we used a combination of 1.4 GHz FIRST radio luminosity and SDSS total spectroscopic star formation rate to separate galaxies into star formation and AGN dominated systems. Figure~\ref{fig:Lradio_excess} shows that this crude diagnostic agrees well with our VLBI results.

\begin{figure}
\includegraphics[width=0.35\textwidth,clip,angle=270]{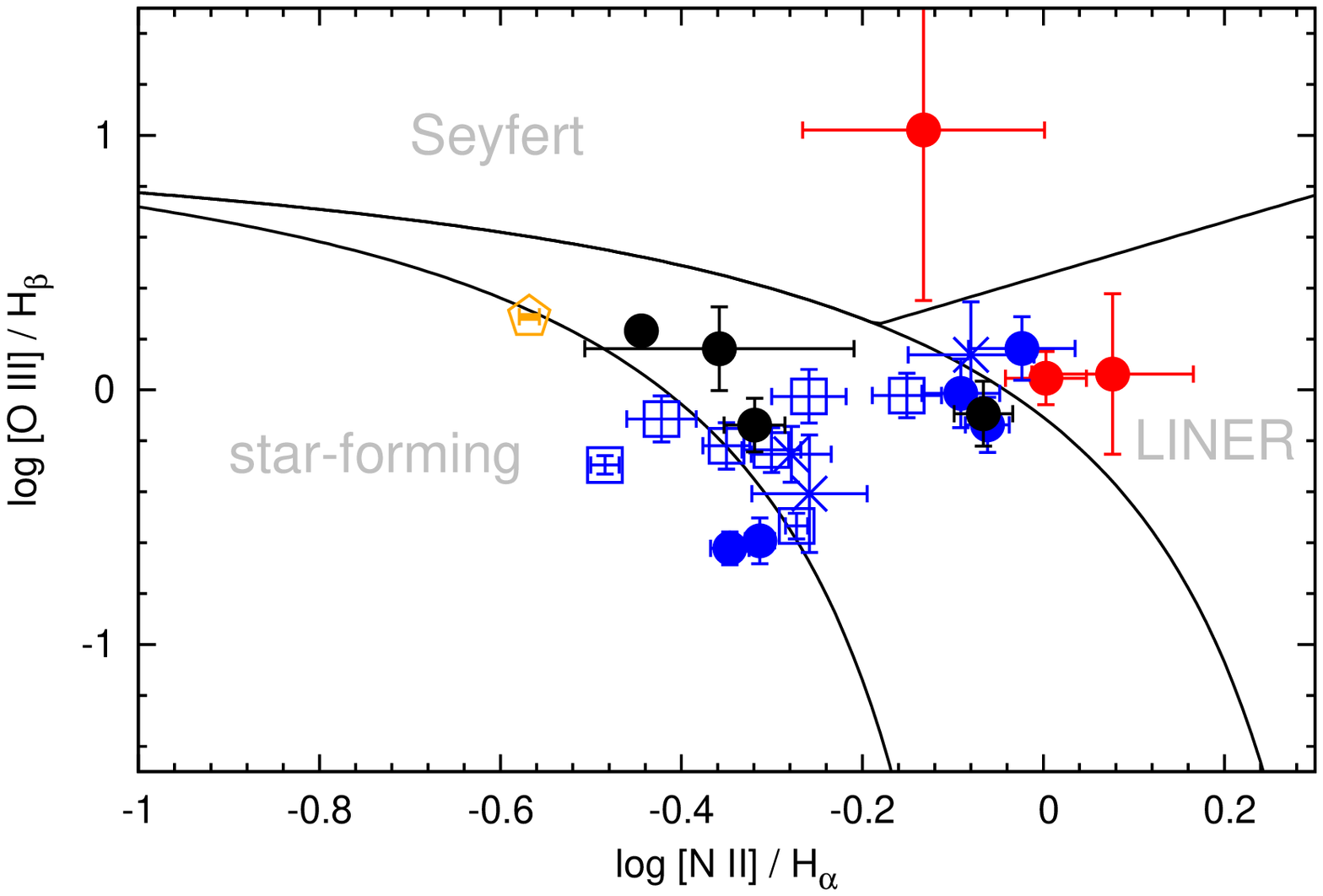}
\includegraphics[width=0.35\textwidth,clip,angle=270]{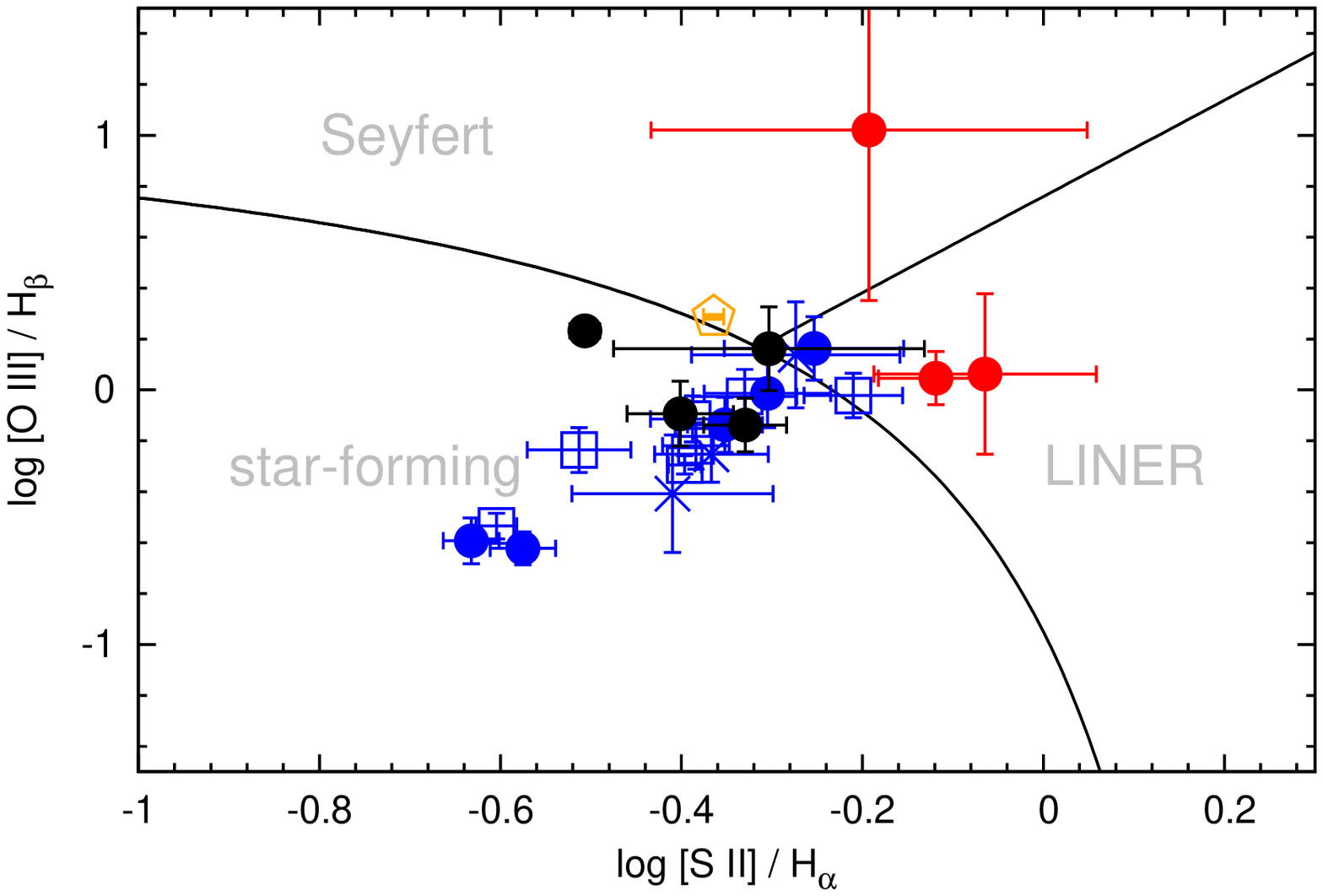}
\includegraphics[width=0.35\textwidth,clip,angle=270]{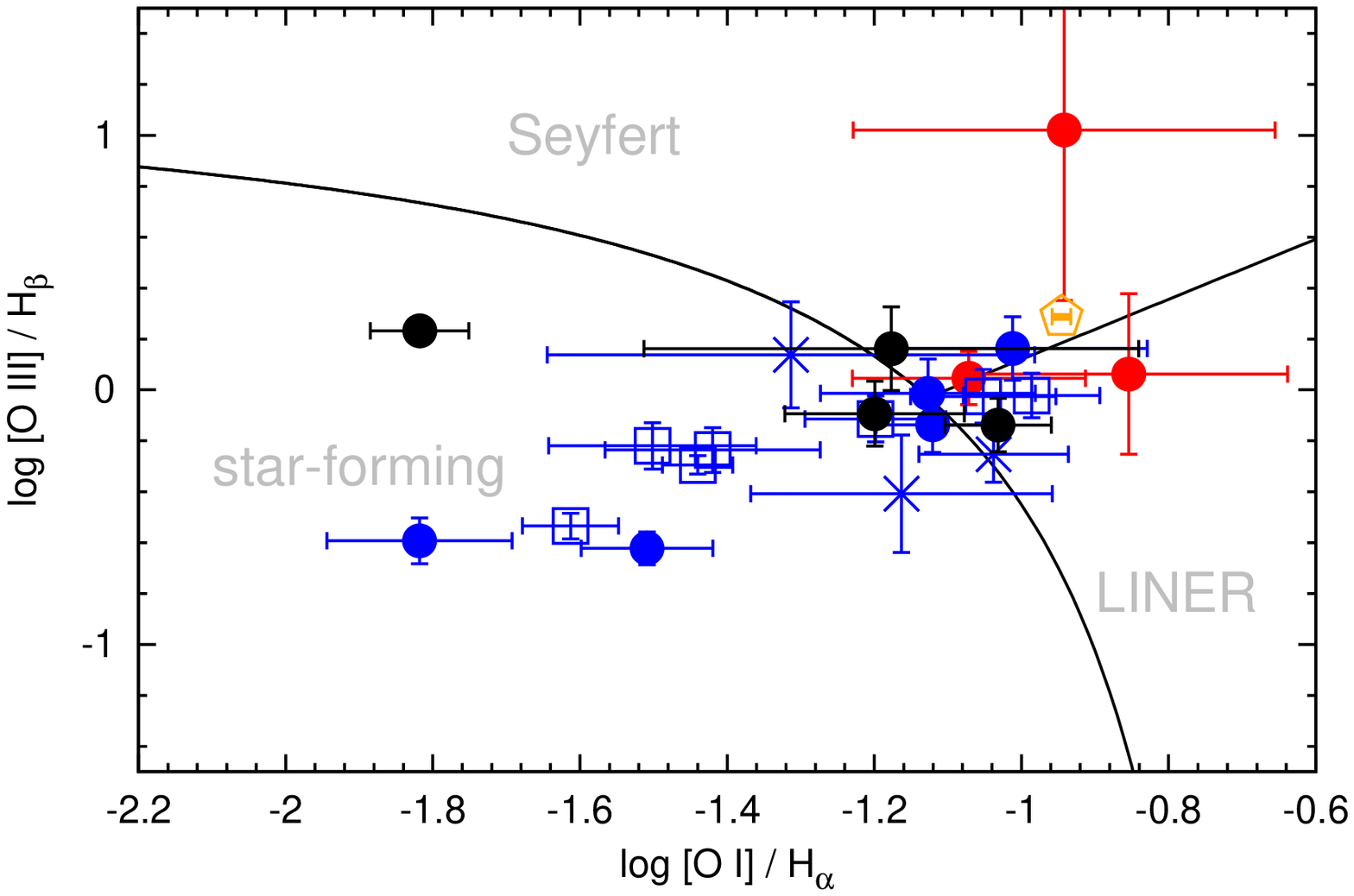}
\caption{Emission line properties of VLBI targets. Symbols are as in Figure~\ref{fig:Lradio_excess}. VLBI observations generally agree well with the [N\,II]$/$H$_\alpha$ ({\it{top panel}}), [S\,II]$/$H$_\alpha$ ({\it{middle}}) and [O\,I]$/$H$_\alpha$ (bottom) optical diagnostics from Kewley et al. (2006). The ``SF dominant'' galaxy in the LINER part of both the [N\,II]$/$H$_\alpha$ and [O\,I]$/$H$_\alpha$ diagnostics is UGC\,7098. The ``SF probable'' galaxy in the LINER part of the [O\,I]$/$H$_\alpha$ and the composite region of the [N\,II]$/$H$_\alpha$ plot is UGC\,10205.}
\label{fig:BPT_mjive}
\end{figure}

Another way of identifying AGN is by using optical emission line ratios. Figure~\ref{fig:BPT_mjive} shows the emission-line diagnostic plots of Kewley et al. (2006) applied to our sample. We find excellent general agreement between our high-resolution radio and emission line properties of the galaxies. Within uncertainties, all VLBI detections lie in the Seyfert / LINER part of the three emission-line diagnostics. Using the [S\,II]$/$H$_\alpha$ ratio (middle panel of Figure~\ref{fig:BPT_mjive}), all VLBI non-detections are consistent with lying on or below the maximum starburst line. In the [N\,II]$/$H$_\alpha$ and [O\,I]$/$H$_\alpha$ diagnostics, there are two galaxies classified as ``SF dominant'' or ``SF probable'' which lie away from the star-forming locus. The galaxy classified as ``SF dominant'' that appears in the LINER part of both the [N\,II]$/$H$_\alpha$ and [O\,I]$/$H$_\alpha$ diagnostics is UGC\,7098. The ``SF probable'' galaxy in the LINER part of the [O\,I]$/$H$_\alpha$ plot also appears in the composite region of the [N\,II]$/$H$_\alpha$ plot; this is UGC\,10205.

In the case of UGC\,7098, the large uncertainties in line flux measurements place this galaxy very close to the demarcation lines. As we discuss below, the mid-infrared $WISE$ colours of this galaxy are consistent with the star forming population. Furthermore, it has extended FIRST radio emission throughout its optical disk; the VLBI fraction of the total radio emission from this galaxy is $<0.17$ and the upper limit on 1.4 GHz AGN luminosity is $4.4 \times 10^{21}$\,W\,Hz$^{-1}$. Hence, it is likely that this galaxy is correctly classified as star-forming. On the other hand, UGC\,10205 has $WISE$ colours that are similar to the VLBI detected AGN, and thus may be mis-classified.


\begin{figure}
\includegraphics[width=0.35\textwidth,clip,angle=270]{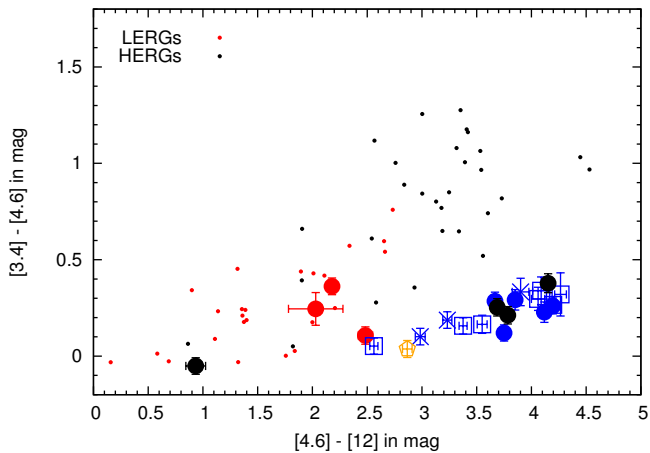}
\caption{$WISE$ colours of VLBI targets. Large symbols are as in Figure~\ref{fig:Lradio_excess}. Small points show local ($z<0.3$) Low and High Excitation Radio Galaxies (LERGs and HERGs, respectively) from the 3CRR and 2 Jansky samples (G\"urkan et al. 2014). The galaxy with $WISE$ colours similar to the AGN population but classified as ``SF probable'' is UGC\,10205. The ``unknown'' galaxy at bottom left is CGCG\,$270-035$.}
\label{fig:WISE_colours}
\end{figure}

We next examine the mid-infrared $WISE$ colours of our galaxies (Wright et al. 2010). The shorter $3.4 \mu$m (W1) and $4.6 \mu$m (W2) $WISE$ bands are dominated by hot dust, which could be heated by either AGN activity or star formation. The $12 \mu$m (W3) band on the other hand is dominated by PAH emission at 11.3$\mu$m and dust continuum, and is indicative of star formation. The mid-infrared colour-colour diagram constructed using these bands is therefore a useful discriminator of passive systems, AGN hosts and star forming galaxies (e.g. Jarrett et al. 2011, Stern et al. 2012). We plot the distribution of $[3.4 - 4.6] \mu$m vs $[4.6 - 12.2] \mu$m colours for our sample in Figure~\ref{fig:WISE_colours}. Both detected and undetected mJIVE-20 targets have similar $[3.4 - 4.6] \mu$m colours, $0 < W1-W2 < 0.5$. However, there is a clear separation in $[3.4 - 12] \mu$m colours, with VLBI detections being significantly bluer ($2.0 < W2-W3 < 2.5$) than all but one of the non-detections ($3.5 < W2-W3 < 4.5$). The location of mJIVE-20 non-detections in the $WISE$ $[3.4 - 4.6] \mu$m vs $[4.6 - 12.2] \mu$m colour space is consistent with optically and radio-identified star forming galaxies (Cluver et al. 2014, Banfield et al. 2015). Our VLBI detections occupy a similar range in this plot to low-redshift ($z<0.3$) Low Excitation Radio Galaxies (LERGs, G\"urkan et al. 2014); we note that this is quite different to the more powerful High Excitation Radio Galaxies (HERGs), which have redder colours consistent with standard $WISE$ AGN classifications (e.g. Stern et al. 2012). We note that Best \& Heckman (2012) classify two of our detections (MJV02278 and MJV16230) as LERGs, with no classification available for MJV16427. The galaxy classified as star forming using VLBI with $WISE$ colours in the AGN part of the diagram is UGC\,10205. As discussed above, it is also the only object classified as star forming based on VLBI observations in Table~\ref{tab:observations} that appears in the LINER part of the [S\,II]$/$H$_\alpha$ and composite part of the [N\,II]$/$H$_\alpha$ vs [O\,III]$/$H$_\beta$ plots of Figure~\ref{fig:BPT_mjive}. It is therefore possible that this object is mis-classified. The ``unknown'' target with unusually blue $WISE$ colours is CGCG\,$270-035$; its mid-infrared colours are consistent with the passive elliptical population (Wright et al. 2010), in line with the low specific star formation rate observed in this galaxy (Table~\ref{tab:observed_sources}).

Interestingly, CGCG\,$270-035$ and UGC\,10205 are the only galaxies in our sample residing in haloes more massive than $3 \times 10^{12} M_\odot$, according to the Yang et al. (2007) group catalogue: UGC\,10205 is part of the Abell\,2162 cluster; CGCG\,$270-035$ is in cluster MSPM\,1032 (Smith et al. 2012). These atypically dense environments (compared with the rest of our sample) may be responsible for their low specific star formation rates and a shift towards the LINER part of line ratio diagnostic plots (Sarzi et al. 2010).

Finally, archival search of X-ray data associated with our targets revealed only one point source within a 15 arcsecond radius of target position. This is IC\,1182, classified as a quasar by V\'eron-Cetty \& V\'eron (2006). This object is also classified as a Seyfert\,2 using [O\,III]$/$H$_\beta$ vs [O\,I]$/$H$_\alpha$ line ratios in Figure~\ref{fig:BPT_mjive}.

Overall, our VLBI AGN classifications agree well with optical narrow line (Figure~\ref{fig:BPT_mjive}), infrared colour (Figure~\ref{fig:WISE_colours}) and X-ray diagnostics, suggesting that none of our VLBI non-detections are likely to be radio-quiet AGN, with the possible exception of UGC\,10205. The quasar IC1182 is classified as ``unknown'' using our VLBI observations. While radio AGN are also always classified as Seyferts or LINERs in our radio-selected sample, the opposite is not true: optical AGN activity can also exist without radio jets in ``composite'' objects, none of which host VLBI sources. We return to this point in Section~\ref{sec:SFH}.

We note that all three of our VLBI-detected AGN are observed to be compact on arcsecond (FIRST) scales, and two are also compact on milli-arcsecond (mJIVE-20) scales at radio wavelengths. The immediate interpretation of this result is that we may be observing these AGN almost immediately after they are triggered, and before they have managed to impart significant feedback on their host galaxies. We return to this point in more detail in Section~\ref{sec:lobeLumins}.

\section{Star formation rates}
\label{sec:SFRs}

The homogeneity of our dust lane sample allows us to study the mechanisms responsible for the triggering of AGN activity in galaxy interactions. We begin by examining the star formation rates for our galaxies, obtained from H$\alpha$ line fluxes in the value-added MPA-JHU spectroscopic catalogues for SDSS Data Release 7 (Brinchmann et al. 2004).

\begin{figure}
\includegraphics[width=0.34\textwidth,clip,angle=270]{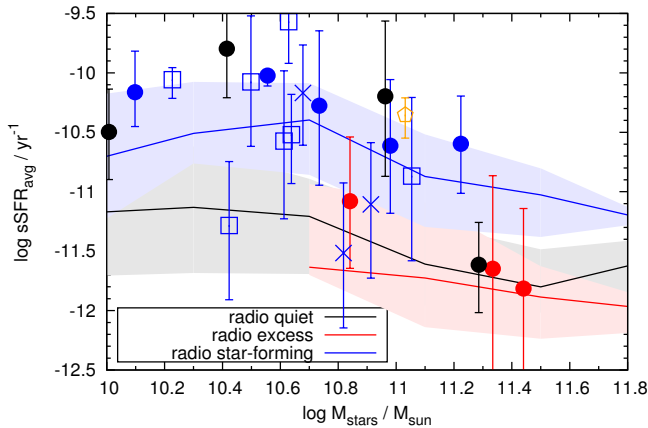}
\caption{Distribution of specific star formation rates as a function of stellar mass. Points are for the mJIVE-20 sample, and shaded regions represent the median and interquartile ranges for the Shabala et al. (2012) sample. Galaxies with excess radio emission (tentatively identified as AGN) have the same sSFR distribution at a given stellar mass as radio-quiet objects; this distribution is also consistent with the sSFR values for the three mJIVE-20 detections presented in this paper. FIRST radio sources consistent with star-forming galaxies have higher sSFRs than both the radio-quiet and radio AGN populations, and these sSFR values are consistent with the mJIVE-20 identified star-forming galaxies. Symbols are as in Figure~\ref{fig:Lradio_excess}. The massive, low sSFR galaxy classified as ``unknown'' is CGCG\,$270-035$.}
\label{fig:sSFR_dist}
\end{figure}

In Figure~\ref{fig:sSFR_dist} we show the specific star formation rates as a function of stellar mass for our VLBI sample. At a given stellar mass, VLBI-detected AGN have lower specific star formation rates than VLBI non-detections. These findings are consistent with the results obtained for the larger sample of Shabala et al. (2012). In that earlier work, objects with radio luminosities in excess of that expected from star formation alone were tentatively classified as AGN. As seen in Figure~\ref{fig:sSFR_dist}, at a given stellar mass radio-quiet galaxies and AGN have the same (confirmed by a Kolmogorov-Smirnov test) low specific star formation rates, while galaxies in which the radio emission can be attributed to star formation have significantly (at the 0.1 percent level) higher specific SFRs.

VLBI observations add an important piece to this puzzle. The original classifications of Shabala et al. (2012) were based on a comparison of radio luminosities and star formation rates. This approach is biased against objects in which the AGN does not completely dominate radio emission -- for example, galaxies in which both AGN and star formation activity are prominent at a similar level. Hence, objects with low star formation rates were more likely to be identified as AGN, while objects with higher SFRs more likely to be classified as star-forming. Our VLBI-selected sample does not suffer from this bias. In other words, if any of our radio AGN co-existed with vigorous star formation, we would be able to classify these objects as such. The observed deficit of significant star formation in radio AGN hosts is therefore real.

Our results add to the large body of existing work on star formation in AGN host galaxies (see e.g. reviews by Alexander \& Hickox 2012, Heckman \& Best 2014). A wide variety of results from multi-wavelength analyses have been reported, ranging from correlations of varying strength between these parameters (e.g. Shao et al. 2010, Santini et al. 2012, G\"urkan et al. 2015), to no correlation at all (e.g. Bongiorno et al. 2012), and even anti-correlations (Page et al. 2012) where AGN feedback is interpreted to suppress star formation. There is a strong dependence on both the cosmic epoch and AGN luminosity, with the star formation -- AGN luminosity correlation being strongest for powerful AGN at $z \gtrsim 1$; a much weaker correlation is found for low-luminosity AGN (Rosario et al. 2012, Rovilos et al. 2012). This is consistent with theoretical work, with galaxy formation simulations predicting a very large scatter (up to 2 orders of magnitude) in star formation rate at a given X-ray AGN luminosity (Sijacki et al. 2015). 

The above studies are typically based on flux-limited samples that cover a range of physical processes responsible for star formation and AGN activity, and probe a wide range in redshift. By contrast, our morphologically-selected sample probes a very specific process (gas-rich minor merging) in the local Universe. We note that the [O\,III] luminosities of our VLBI detections are in the range $10^6 - 2 \times 10^7$\,L$_\odot$, consistent with the ``weak AGN'' classification of Kauffmann et al. (2003) and Wild et al (2010). Kauffmann et al. (2003) found such AGN to have stellar populations that are similar to the overall early-type population (see their Figure 13). These AGN and their hosts are likely a very different population to, for example, those found in X-ray selected samples, which probe powerful AGN in a much younger Universe. The different accretion histories and luminosities of the AGN making up the two types of samples are likely responsible for the very different observed relationships between AGN activity and star formation. This is an important consideration even at low redshift: for example, radio AGN are known to be more prevalent at the centres of clusters with short cooling times (Mittal et al. 2009); these clusters also have bluer central galaxies (Rafferty et al. 2006). Averaging over all environments (i.e. cool core and non-cool core clusters) will therefore yield a correlation between the presence of an AGN and star formation rate. By only focusing on gas-rich mergers, we attempt to account for such environmental effects with our present sample.

\section{Star formation histories}
\label{sec:SFH}

In this section, we use optical (SDSS; Abazajian et al. 2009) and ultraviolet (GALEX; Martin et al. 2005) photometry to reconstruct the star formation histories of our galaxies. This combination is particularly powerful at quantifying recent ($\sim 2$~Gyr) starburst activity, even if the mass fraction of newly formed stars is relatively low (e.g. Kaviraj et al. 2007).

\subsection{Stellar population fitting}
\label{sec:SFH_fitting}

Stellar population histories were reconstructed by comparing GALEX ({\it{NUV}}) and SDSS ({\it{ugriz}}) photometry with a large library of synthetic photometry computed for a wide range of modelled star formation histories. These star formation histories are tailored to early-type galaxies which dominate our sample (Kaviraj et al. 2012, Shabala et al. 2012). Because the bulk of the stellar mass in early-type galaxies is assembled at high redshift, we model this old stellar population with an instantaneous burst at $z=3$. The recent, merger-triggered, star formation episode is represented as a second instantaneous burst the age of which is allowed to vary between 1~Myr and 10~Gyrs. The mass fraction of this burst is allowed to vary between 0 and 1. Model star formation histories are combined with metallicities in the range $0.1-2.5$Z$_\odot$, and dust extinction is parametrized following Calzetti et al. (2000) using $E_{\rm B-V}$ values in the range $0-0.5$. Each model star formation history is convolved with the stellar models of Yi (2003) to yield the predicted GALEX and UV photometry. We construct such model photometries in redshift steps of $\delta z=0.02$. 

We apply these models to the observed photometry in our sample by comparing each galaxy to every model in the synthetic library. Model likelihoods\footnote{Likelihood is proportional to exp$(-\chi^2/2)$.} are calculated for each model--galaxy combination, and probability density functions are constructed for the age of the most recent starburst, initial merger gas fraction, metallicity and dust extinction, by marginalising over the joint probability distributions. The medians of these probability density functions are adopted as the best estimates for each parameter, and the 16th and 84th percentile values as $1\sigma$ uncertainties. In the following sections we use the starburst ages estimated using this method. These are given in column (5) of Table~\ref{tab:observed_sources}.

\subsection{Evolution of the star formation rate}
\label{sec:SFR_evolution}

\begin{figure}
\includegraphics[width=0.33\textwidth,clip,angle=270]{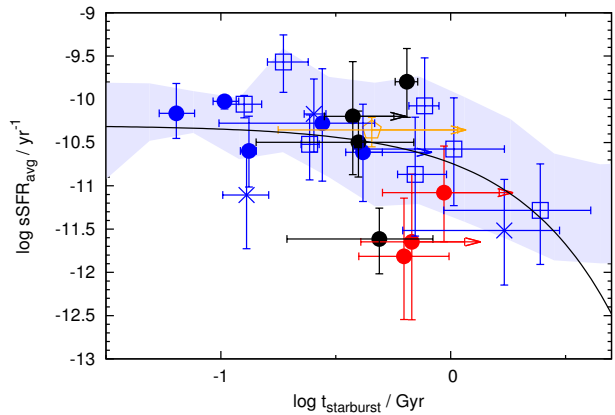}
\caption{Specific star formation rate as a function of stellar age. Shaded region shows the $1\sigma$ confidence interval from Monte Carlo sampling of sources classified as ``SF dominant'' and ``SF probable''. Solid line shows an exponential evolutionary track with an initial gas fraction of 5 percent, and a decay timescale of 1~Gyr. Older galaxies appear to have lower sSFRs, consistent with depletion of the gas supply brought in during the minor merger. There are no AGN younger than 400 Myr, and AGN appear only when the star formation rate drops to low values. The ``unknown'' galaxy with an old starburst age and low sSFR is CGCG\,$270-035$; the ``SF plausible'' galaxy with an old starburst is MCG$-116$.}
\label{fig:SFR_tAge}
\end{figure}

Assuming that our 23 dust lane galaxies with star formation histories are drawn from a homogeneous sample, and we simply happen to observe them at different evolutionary stages, the derived starburst ages can be treated as merger clocks. In Figure~\ref{fig:SFR_tAge} we examine the evolution of the specific star formation rate with time since the onset of the starburst. Upon first inspection, galaxies with older starbursts appear to have lower specific SFRs, as expected if the gas supply brought in during the merger is continuously depleted by star formation. The large scatter in this apparent trend is due to a number of factors. First, both the star formation rate and starburst age show large formal uncertainties (typically 0.5~dex). Second, merger-triggered star formation is expected to be ``bursty'', with episodes of high star formation activity corresponding to pericentric passages of the gas-rich satellite (e.g. Peirani et al. 2010). Furthermore, the gas mass brought in by the merging satellite sets the normalisation of the star formation rate. On average, however, depletion of the gas reservoir is expected to yield an exponentially-decaying rate of star formation with a timescale that depends on details such as the mass ratio of the merging galaxies. In Figure~\ref{fig:SFR_tAge} we show that a putative gas-rich merger with an initial molecular gas fraction of 5 percent (the median value inferred from CO observations of a subsample of our dust lane galaxies; Davis et al. 2015) and a decay timescale of 1~Gyr is consistent with the data.

We test the statistical significance of the specific star formation -- age relation using Monte Carlo simulations with replacement. Here, we performed 30,000 bootstrap resampling realizations of our data, using reported formal uncertainties in both star formation rate (Brinchmann et al. 2004) and starburst ages (from photometric fitting). The resultant non-parametric $1\sigma$ uncertainties are plotted as the shaded region in Figure~\ref{fig:SFR_tAge}. The apparent decrease in star formation rate with age is not significant at the $2\sigma$ level, however the uncertainties in the two quantities for individual galaxies are quite large. Better measurement of the stellar ages and star formation rates, and quantification of the molecular gas fraction associated with the merger (through CO observations) are required to properly test this relation.


Regardless of the above discussion, Figure~\ref{fig:SFR_tAge} shows that the radio AGN are only triggered when the starburst age exceeds at least 400~Myr (this value corresponds to using the large uncertainties on starburst age; the best-fit age for the youngest VLBI-detected AGN is 600~Myrs), and/or the specific SFR drops below $3 \times 10^{-11}$\,M$_\odot$\,year$^{-1}$. These limits apply even if CGCG\,270-035 (classified as ``unknown'' using VLBI data) and UGC\,10205 (``SF probable'') are classified as AGN.

Selection effects must be considered when interpreting these findings. One possibility for the association between VLBI-detected AGN and low specific SFRs is that this is simply a selection effect, and the radio AGN in galaxies with more vigorous star formation may be present but are simply not detectable. This is unlikely to be the case, however. All three of our AGN detections have sufficiently high radio luminosities (Table~\ref{tab:observations}) to make them detectable in all galaxies in our sample classified as star formation dominant or probable, including objects with high star formation rates. Thus, the lack of radio AGN activity in young post-starburst galaxies with vigorous star formation activity is real.

The VLBI detections in our sample agree well with optical AGN diagnostics, with all VLBI AGN lying in the Seyferts or LINER part of the emission line diagnostic plots, and no galaxy clearly classified as a Seyfert or LINER is ``star formation dominant'' or ``star formation probable'' in our VLBI classification. This is not a radio selection effect: Shabala et al. (2012) studied a much larger sample of 484 dust lane galaxies, and found no difference in the stellar population ages between galaxies with and without radio emission. Regardless of the presence of radio emission, these authors found a trend towards increasing starburst age from objects classified using emission lines as star forming, to composite objects, to Seyferts and LINERs, consistent with previous findings of Schawinski et al. (2007) and Wild et al. (2010). The derived starburst ages for each class did not change depending on whether the dust lane galaxy hosted a radio AGN (identified using excess in FIRST emission compared to the H$\alpha$-derived star formation rate); those authors did, however, find a strong correspondence between radio and Seyfert/LINER emission line AGN activity. The present work is consistent with those findings.

The correspondence between Seyfert/LINER and radio AGN classifications suggest that the radio AGN are triggered relatively late during the merger stage. If galaxies classified as ``composite'' objects in the BPT emission line diagnostic are interpreted as hosting both optical AGN and star formation activity, the lack of VLBI detections in these objects provide further support for the evolutionary sequence proposed by Cowley et al. (2016; see also Schawinski et al. 2015), in which radio AGN represent a more evolved post-merger stage than either X-ray, infra-red or optically-selected AGN.

The causal connection between radio AGN activity and low star formation rates is unclear. On the one hand, jet generation models predict higher jet production efficiencies at low accretion rates (e.g. Meier 2001, Benson \& Babul 2009). A common mechanism such as exhaustion of the gas reservoir may facilitate both cessation of star formation and jet generation (e.g. Norman \& Scoville 1988, Davies et al. 2007, Wild et al. 2010); in other words, the radio jets may be switching on because both the gas supply and star formation have been sufficiently depleted. Alternatively, the observed low levels of star formation could be due to AGN feedback imparted by the active nucleus.

\section{Feeding or feedback?}
\label{sec:lobeLumins}

Radio AGN observables such as size, luminosity and spectral index encode much important information related to AGN physical parameters. Dynamical models of radio AGN (Kaiser et al. 1997, Blundell et al. 1999, Shabala et al. 2008, Turner \& Shabala 2015) and/or spectral ageing techniques (Alexander \& Leahy 1987, Murgia et al. 1999) allow for an inversion of these observables to meaningful quantities such as AGN ages and jet kinetic powers. All three VLBI-detected AGN in our sample appear compact on kpc-scales in the FIRST and NVSS surveys, placing an upper limit of $\sim 5$~arcsec on their size. This corresponds to 13 kpc for our most distant AGN (MJV02278),  and 2 kpc for the nearest detection (MJV16427). With typical lobe expansion speeds of $0.01 - 0.1 c$ (Alexander \& Leahy 1987, de Vries et al. 2009) this translates to ages younger than 5 Myrs, which is significantly less than the hundreds of Myrs ages of the largest AGN (Blundell \& Rawlings 2000, Shabala et al. 2008). It is puzzling that no large (tens to hundreds of kpc) radio AGN are observed either in the present VLBI sample, or the radio-excess AGN sample of Shabala et al. (2012).

One tantalising possibility is that the observed AGN sizes do not provide an accurate indication of the AGN's true extent. This would happen if, for example, the surface brightness of the synchrotron-emitting radio lobes falls below the survey detection sensitivity (e.g. Sadler et al. 1989). In this case, only the compact self- or free-free absorbed core (but not the lobes) will be detected, and the AGN will appear compact.

We test this scenario by modeling the temporal evolution of the lobe surface brightness for each of our three VLBI detections. To this end, we use the RAiSE (Radio AGN in Semi-analytic Environments) model of Turner \& Shabala (2015). This package employs semi-analytic galaxy formation models to quantify the gaseous atmospheres into which the AGN are expanding, and can describe both high and low luminosity radio AGN  populations. The required inputs for this model are the AGN jet kinetic power and some measure of the AGN environment, typically either the stellar or dark matter halo mass of the AGN host. We calculate the jet kinetic power following the prescriptions of Meier (2001) for two accretion scenarios: a standard Shakura-Sunyaev thin disk, and an Advection Dominated Accretion Flow,

\begin{eqnarray}
Q_{\rm jet, TD} = 5 \times 10^{37} k_{\rm TD} \left( \frac{M_{\rm BH}}{10^9} \right)^{0.9} \dot{m}_{\rm BH}^{1.2} \, {\rm W} \nonumber\\
Q_{\rm jet, ADAF} = 6.7 \times 10^{38} k_{\rm ADAF} \left( \frac{M_{\rm BH}}{10^9} \right) \dot{m}_{\rm BH} \, {\rm W}
\end{eqnarray}
Here, the black hole mass $M_{\rm BH}$ is estimated from the stellar velocity dispersion (Tremaine et al. 2002); $k_{\rm TD}=1-2.4$ and $k_{\rm ADAF} =0.6 - 3$ are parameters related to black hole spin, with the lower values corresponding to non-spinning and higher values to maximally spinning black holes; and the Eddington-scaled accretion rate $\dot{m}_{\rm BH}$ is calculated from the [O\,III] line luminosity, $\dot{m}_{\rm BH} \approx \frac{3500 L_{\rm [O\,III]}}{L_{\rm Edd}}$ (Heckman et al. 2004). Above some critical accretion rate $\dot{m}_{\rm crit}$ the accretion disk is expected to be in a geometrically thin (Shakura-Sunyaev) configuration, while at lower accretion rates it ``puffs up'' and becomes radiatively inefficient (i.e. an ADAF). Observationally, these states are loosely related to High and Low Excitation Radio Galaxies, HERGs and LERGs respectively (Hardcastle et al. 2007). Modeling and observations of state transitions in X-ray binaries suggest that $\dot{m}_{\rm crit} \approx 0.001 - 0.1$ with a large uncertainty in this value (see Meier 2001 and references therein); similar values are obtained through analysis of AGN and galaxy populations (Merloni \& Heinz 2008, Shabala \& Alexander 2009). Using [O\,III] line luminosities, we estimate $\dot{m}_{\rm BH}$ of between $0.0009$ and $0.03$ for our three VLBI-detected AGN. As mentioned in Section~\ref{sec:AGN_comparison}, Best \& Heckman (2012) classify MJV02278 and MJV16230 as Low Excitation Radio Galaxies, suggesting that these may be ADAFs; no classification is given for MJV16427. We note that, due to possible contributions from star formation-driven shocks to the [O\,III] line luminosity, our assumed accretion rates are strictly speaking upper limits.

\begin{figure*}
\includegraphics[width=0.3\textwidth,clip,angle=270]{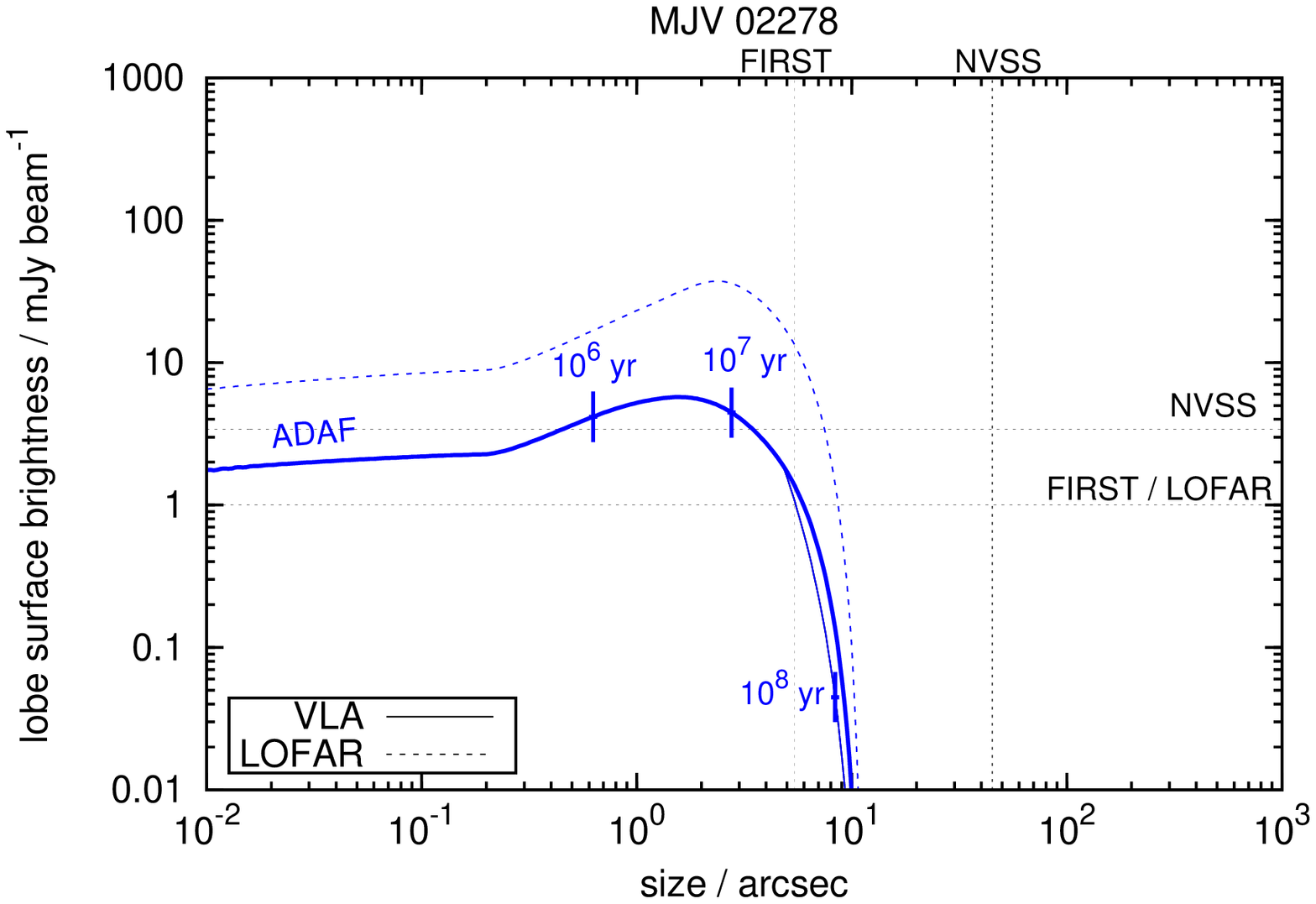}
\includegraphics[width=0.3\textwidth,clip,angle=270]{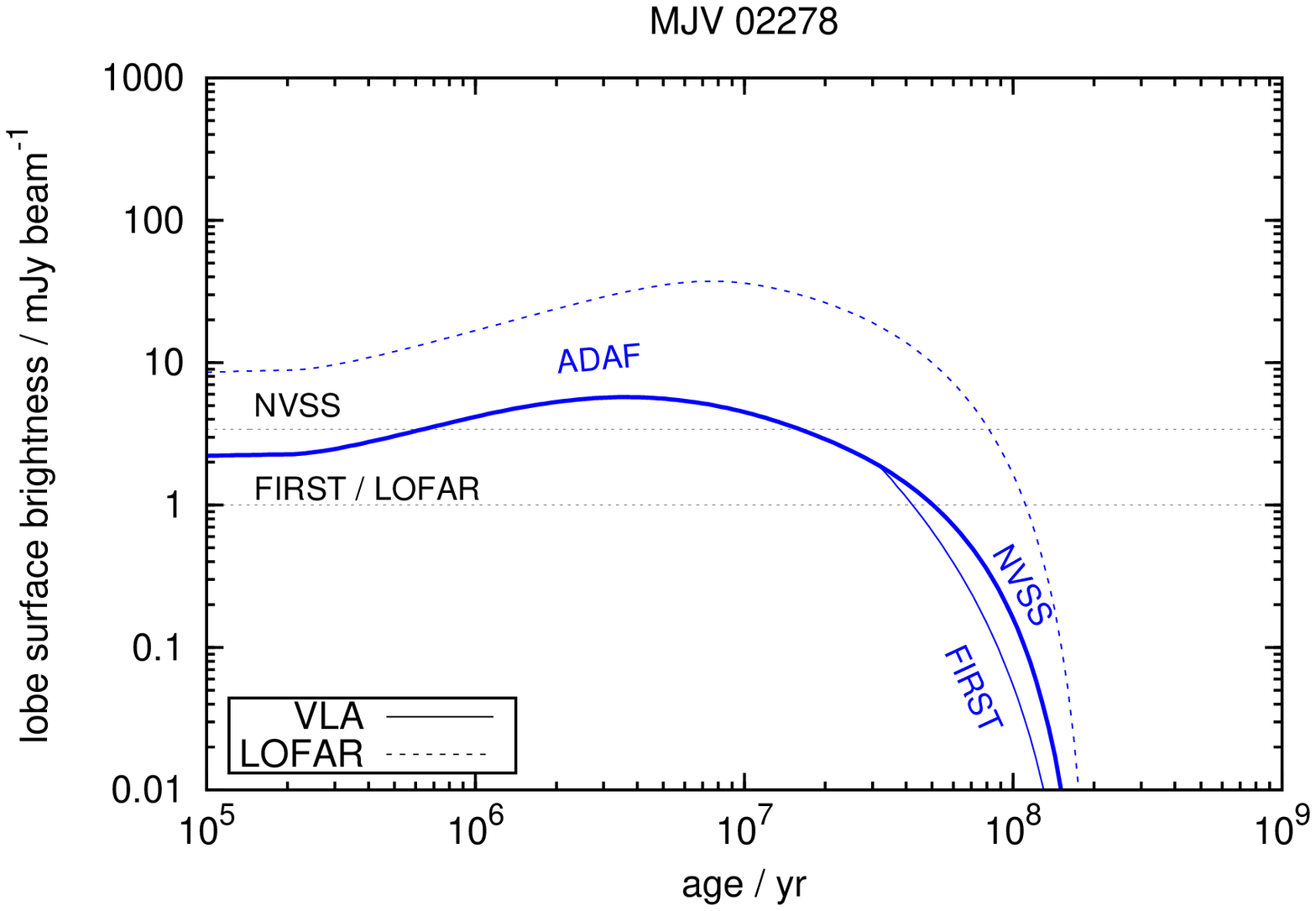}
\includegraphics[width=0.3\textwidth,clip,angle=270]{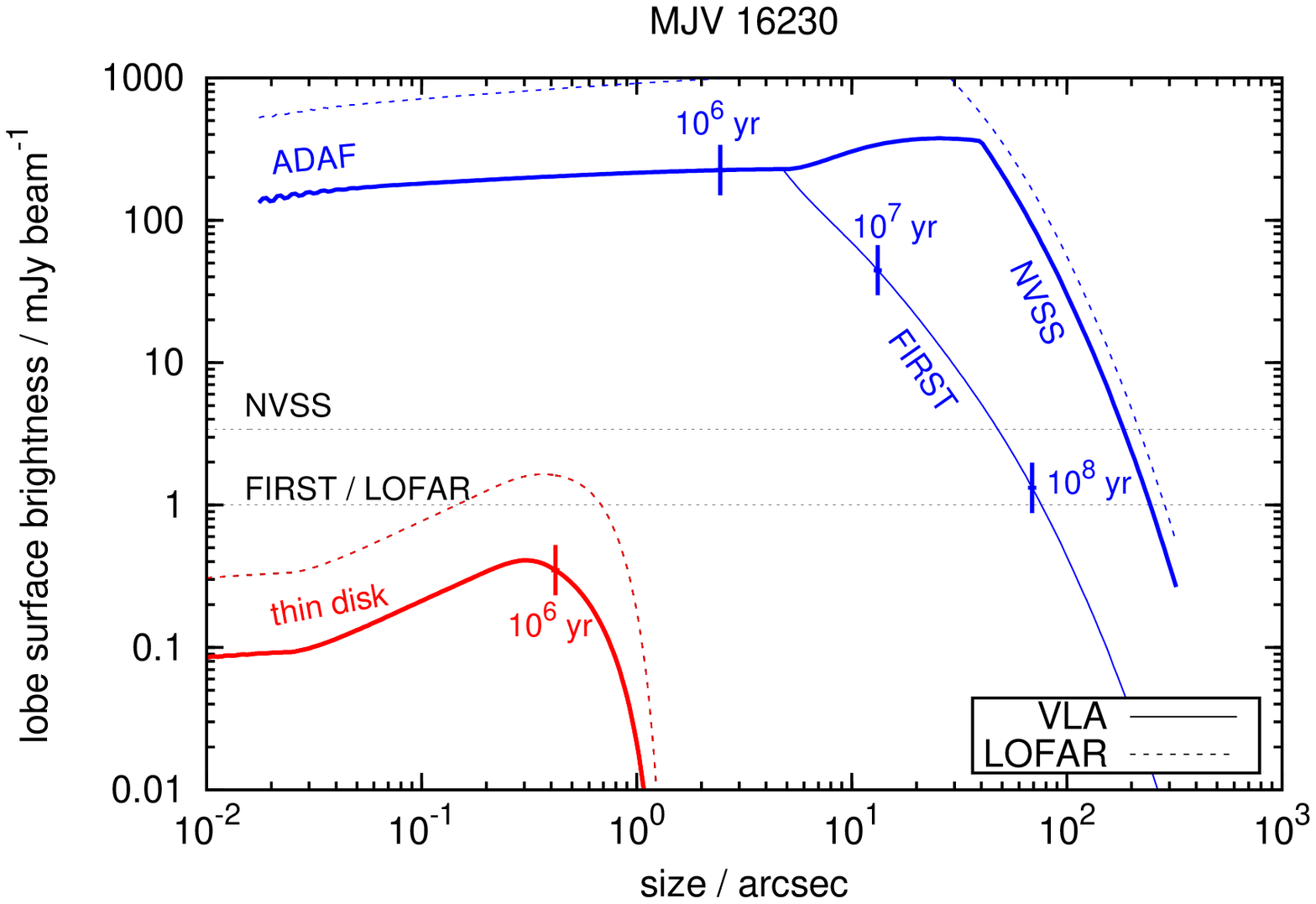}
\includegraphics[width=0.3\textwidth,clip,angle=270]{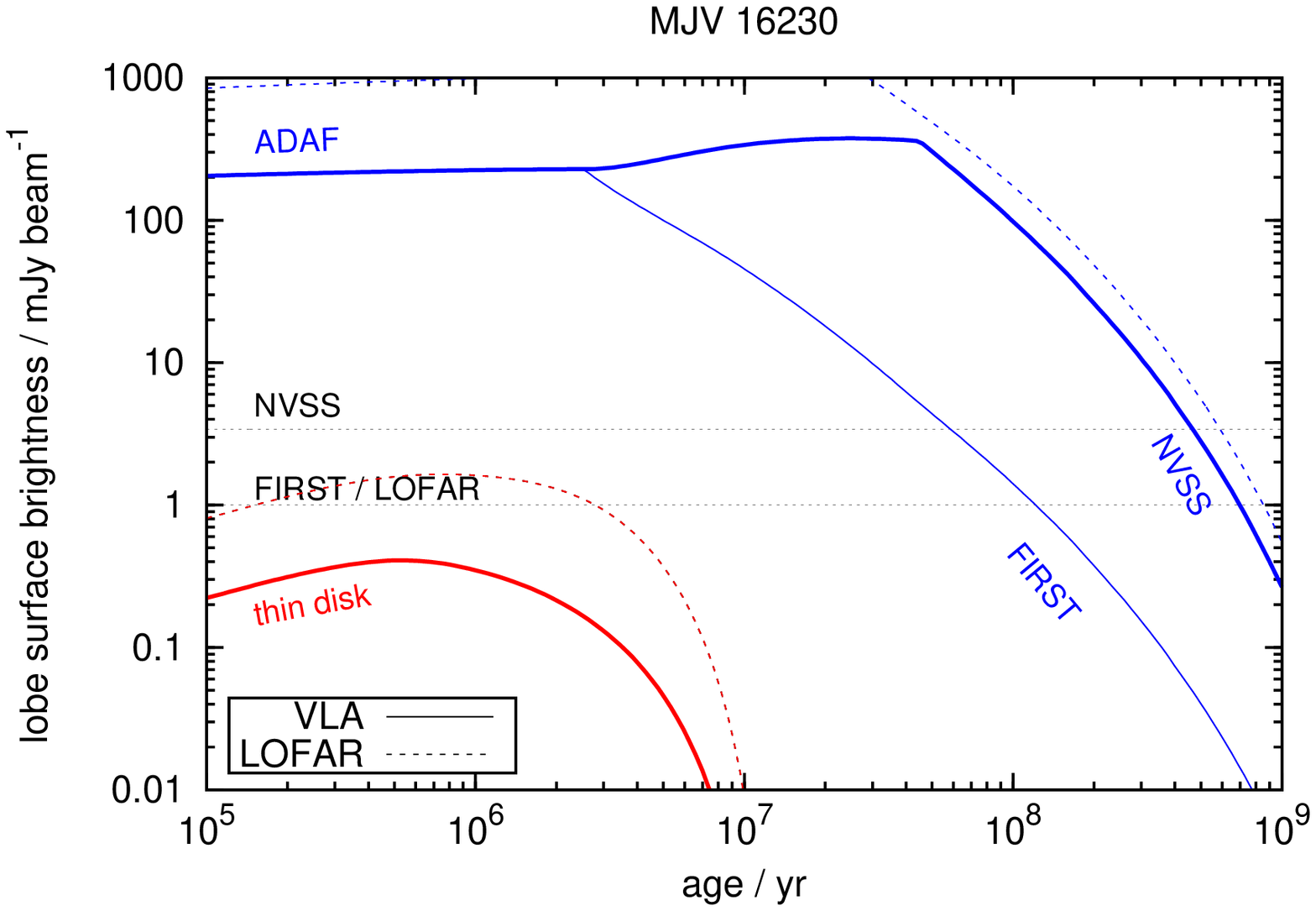}
\includegraphics[width=0.3\textwidth,clip,angle=270]{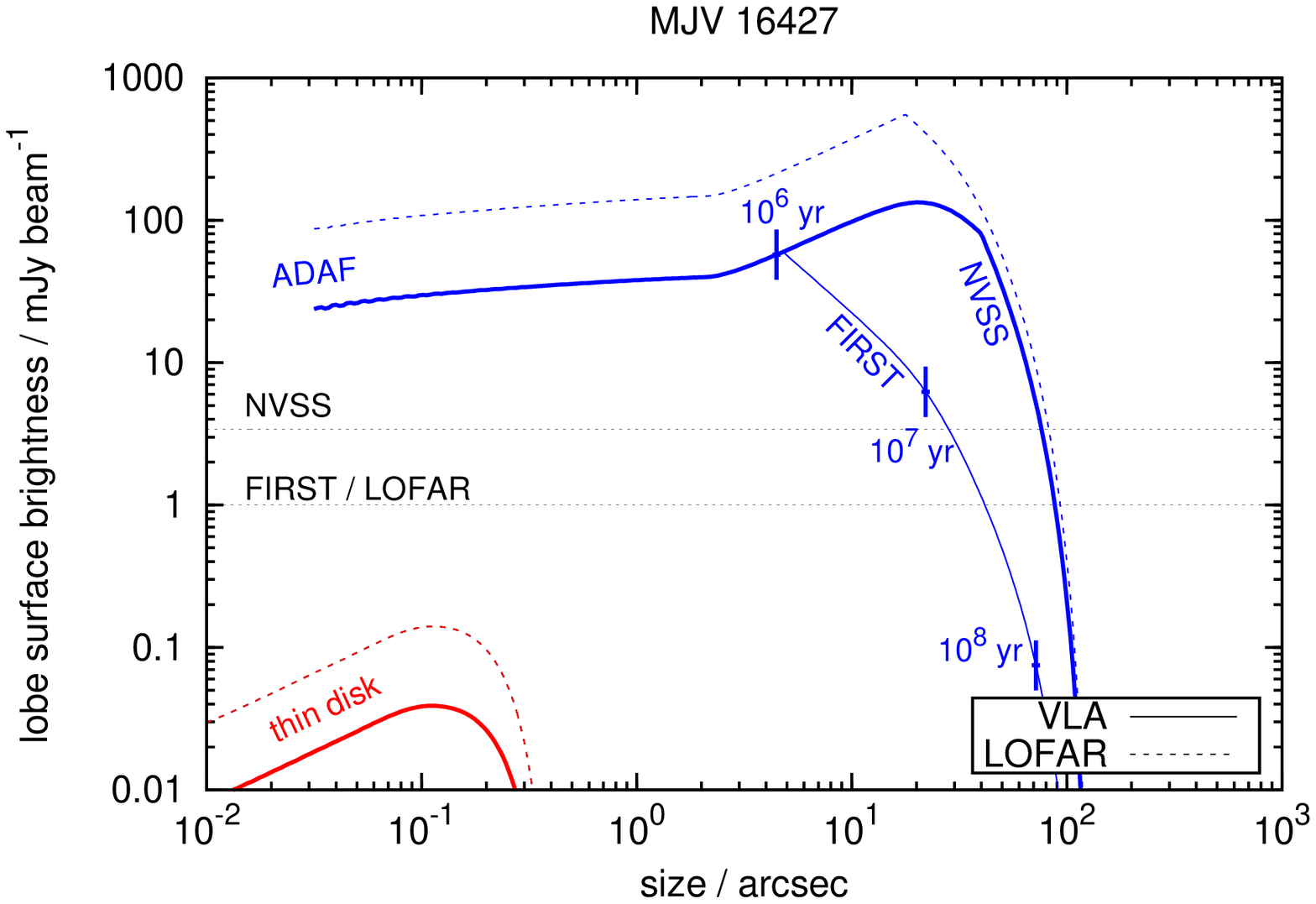}
\includegraphics[width=0.3\textwidth,clip,angle=270]{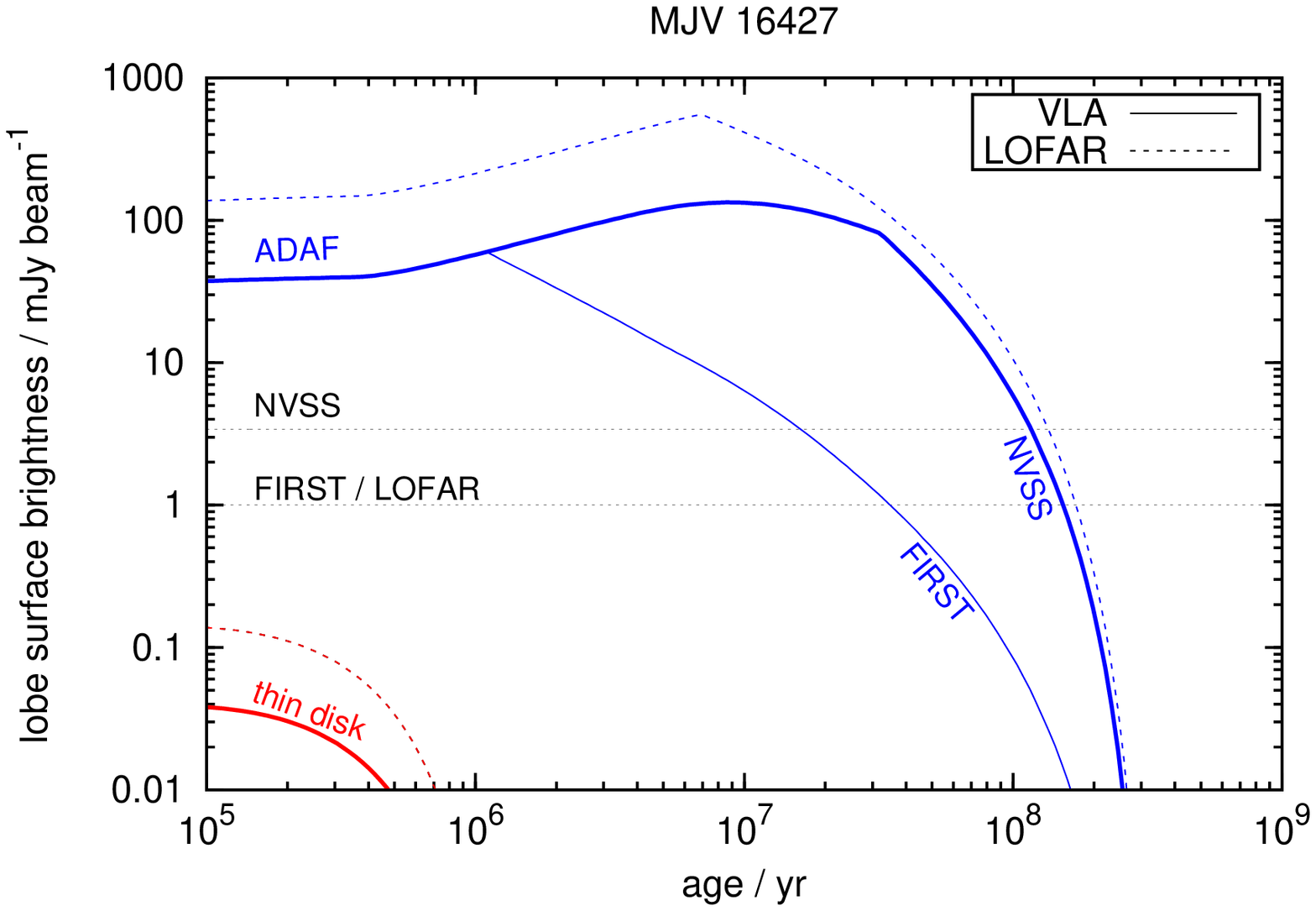}
\caption{Model lobe surface brightness as a function of size (left panel) and age (right panel). Evolutionary tracks are shown for the FIRST and NVSS surveys at 1.4~GHz (solid coloured lines) and LOFAR at 150 MHz (dashed coloured lines). MJV\,02278 is likely powered by an advection-dominated accretion flow (blue lines) while jets in MJV\,16230 are more likely to be produced by a standard optically thin disk (red line). MJV\,16427 is likely to be an intermediate case (see text). Dashed black lines denote survey detection limits. Vertical markers in the left panels indicate source size and surface brightness at various stages of evolution.}
\label{fig:lobeLumin}
\end{figure*}

Figure~\ref{fig:lobeLumin} shows the modelled evolution of lobe surface brightness with size (left) and AGN age (right panel) for both ADAF and thin disk jets.

We investigate the expected lobe surface brightness for a range of existing and upcoming surveys. For the VLA FIRST survey at 1.4~GHz (Becker et al. 1995) we assume 5.4~arcsec beam FWHM and a 1~mJy/beam detection threshold (approximately corresponding to a 6$\sigma$ detection). For the 1.4~GHz VLA NVSS survey (Condon et al. 1998)  we assume 45 arcsec resolution and a 3.4~mJy/beam detection threshold (corresponding to 99 percent survey completeness). Finally, we also investigate the detectability of these lobes at 150~MHz with the Low Frequency Array (LOFAR; van Haarlem et al. 2013), for which we assume 20~arcsecond resolution and a 5$\sigma$ detection threshold of 1~mJy/beam. We note that the uncertainty in our surface brightness predictions at a given age are likely to be a factor of a few (Turner \& Shabala 2015) due to the uncertainty in both AGN jet power and environments into which the radio jets are expanding.

For all three VLBI-detected AGN, the lobes inflated by thin disk jets are close to or below the FIRST and NVSS detection thresholds at any point in their evolution. This model is likely applicable for MJV16230, which has $\dot{m}_{\rm BH}=0.03$. Figure~\ref{fig:lobeLumin} suggests that even LOFAR observations would struggle to detect lobes of this source in the thin disk state.

For the remaining two AGN, the dimensionless accretion rates are 0.0009 (MJV02278) and 0.003 (MJV16427). The first of these is most likely powered by an ADAF, while the latter may be in an intermediate accretion disk state.

The top right panel of Figure~\ref{fig:lobeLumin} shows that, in the case of MJV02278, even ADAF jets cannot produce lobes that remain detectable at 1.4~GHz for longer than about 10~Myrs. On the other hand, sensitive low frequency observations should be able to detect the extended lobes until they are approximately 100~Myrs old; this source would be a good target for LOFAR.

The lack of extended emission in NVSS observations around MJV16427 suggests that this source is most likely not powered by an ADAF, consistent with its ``intermediate'' accretion rate. At a given age, the predicted LOFAR surface brightness is comparable to that for NVSS in both the thin disk and ADAF accretion states (see bottom panels of Figure~\ref{fig:lobeLumin}), and we therefore do not expect to be able to detect the lobes at 150~MHz.

An alternative possibility to the lobes being large and faint is that the true sizes of the AGN lobes are somewhere between the minimum angular scale probed by FIRST (5.4~arcsec) and the largest angular scale probed by mJIVE-20 ($\sim 0.3$~arcsec). In this case, the lobes would be invisible to mJIVE-20 because the available observations are simply not sensitive to the relevant spatial scales. Left panels of Figure~\ref{fig:lobeLumin} suggest that only MJV02278 may fall in this category, and even then only for a source age less than 40~Myr.

In summary, it is highly likely that our three VLBI AGN are not in fact young but simply appear compact due to the limited sensitivity of available observations. We note that this finding very likely has implications beyond our dust lane mJIVE-20 sample. For example, the unexpectedly high observed fraction of compact, low-luminosity AGN in broader FIRST and NVSS samples (Shabala et al. 2008) may be due to this effect. We defer a detailed investigation of this point to a future paper.

For the purposes of the present investigation, it is sufficient to note that the observed spatial extent of the radio AGN in our sample only provides a (not very useful) lower limit on the AGN age. This means that we cannot rule out AGN feedback as the cause of the low levels of star formation associated with our radio AGN. We note that this does not affect our conclusion about the existence of a time delay between the onset of star formation and radio AGN activity, since no galaxies in our sample with stellar populations younger than $\sim 400$~Myrs are found to host radio AGN. By mapping out the extent of both star formation and AGN activity, sensitive spatially resolved radio imaging and Integral Field Spectroscopy would help to test the contribution of AGN feedback to the low observed star formation rates. The ultimate goal is to find genuinely compact AGN which have not yet had time to impart large-scale feedback on their host galaxies. Such objects would allow us to study conditions at the centres of galaxies just as the AGN are triggered.

\section{Efficiency of AGN feedback}
\label{sec:effFeedback}

Although we cannot ascertain the cause of the low star formation rates in our AGN hosts, there exists a clear delay between the onset of merger-triggered star formation and radio AGN activity in our sample. The existence of this delay has important implications for the efficiency of AGN feedback. Figure~\ref{fig:SFR_tAge} suggests that radio AGN do not switch on until at least 400~Myrs, and possibly as long as 600~Myrs, after the onset of merger-triggered star formation, while a plausible {\it e}-folding timescale for the star formation is 1~Gyr. This means that between a third and half of the gas brought in by the merger is already turned into stars\footnote{Mass fraction of newly formed stars is given by $(1-e^{t/\tau})$, where $t$ is the age of the merger-triggered starburst and $\tau$ is the depletion timescale. For $\tau=1$~Gyr and $t=400-600$~Myr, this mass fraction is in the range 0.33 -- 0.45.} before the radio jets switch on. The radio AGN can therefore only couple to the residual gas in the host galaxy, making the resultant feedback much less efficient than if the AGN triggering had been prompt. In a related recent work (Kaviraj et al. 2015a) we found that galaxies hosting recent merger-triggered low-redshift radio AGN preferentially reside on the red optical sequence, confirming that these AGN are unable to significantly suppress star formation once it begins. This underscores the important role at low redshift of hot-mode AGN feedback (Shabala \& Alexander 2009, McNamara \& Nulsen 2012; also known as ``radio mode" feedback, Croton et al. 2006). In this mode, radio AGN operate as thermostats, with gravitational instabilities associated with hot gas cooling triggering AGN activity (Best et al. 2005, Pope et al. 2012) and preventing runaway gas cooling onto galaxies.

\section{Summary}
\label{sec:conclusions}

We presented a morphologically-selected sample of 25 recent gas-rich minor mergers with arcsecond-scale radio emission, observed with Very Long Baseline Interferometry (VLBI). The VLBI technique provides an unambiguous way of identifying radio AGN, including in objects where AGN and vigorous star formation co-exist. Three objects in our sample have VLBI detections. Upper limits on VLBI flux densities and radio morphology allow us to classify a further 12 objects as likely star-forming galaxies, and another three as candidate star-forming galaxies. We find that for our objects the VLBI AGN classification is generally consistent with Seyfert or LINER classification in standard optical emission line diagnostics; none of our VLBI detection AGN are in the ``composite'' part of the BPT diagram. The VLBI AGN classifications also agree well with mid-infrared colour diagnostics, as well as identification of radio AGN via excess luminosity over that expected from the star formation alone. The radio morphologies of our AGN appear compact on arcsecond scales.

We used optical and UV photometry to reconstruct star formation histories for our galaxies, and found that the evolution of specific star formation rates with starburst age in non-radio AGN hosts are broadly consistent with expectations from gas depletion models. The VLBI-identified AGN are triggered no earlier than 400~Myrs after the onset of star formation, severely limiting the efficiency of any feedback these AGN can impart on their host galaxies.

Although our radio AGN appear compact, dynamical modeling of expanding radio lobes shows that our observations may simply have insufficient sensitivity to detect extended radio structures. We therefore cannot rule out AGN feedback as the cause of the low star formation rates in radio AGN hosts. If the observed AGN are genuinely compact, these low star formation rates are likely to be an essential ingredient for AGN triggering, in line with models of stellar feedback limited black hole accretion. On the other hand, if the AGN are extended, the low star formation rates may simply be due to suppression of star formation by AGN feedback. Sensitive, spatially resolved radio and IFU observations will distinguish between these scenarios.

\section*{Acknowledgements}

We thank Elaine Sadler for illuminating discussions. SSS thanks the Australian Research Council for an Early Career Fellowship (DE130101399). ATD was supported by an NWO Veni Fellowship. RJT thanks the University of Tasmania (UTAS) for an Elite Research Scholarship. SK is grateful for support from UTAS via a UTAS Visiting Scholarship and acknowledges a Senior Research Fellowship from Worcester College Oxford. We thank the anonymous referee for a thoughtful report that has helped improve the paper.


\label{lastpage}

\end{document}